\documentclass[9pt,shortpaper,twoside,web]{ieeecolor}
\usepackage{apg}
\makeatletter\let\NAT@parse\undefined\makeatother %correct the fake natbib commands hack found
\usepackage[numbers]{natbib}
\usepackage{amsmath,amssymb,amsfonts,placeins}
\usepackage{algorithmic}
\usepackage{graphicx}
\usepackage{textcomp}
\usepackage{hyperref}
\usepackage{mathtools}%for \mathclap
\usepackage{url} 

\def\BibTeX{{\rm B\kern-.05em{\sc i\kern-.025em b}\kern-.08em
    T\kern-.1667em\lower.7ex\hbox{E}\kern-.125emX}}
\usepackage[switch]{lineno}
\usepackage{subcaption}
\usepackage{booktabs,multirow,accents}

\usepackage{bm}\renewcommand{\vec}{\bm}
  
\makeatletter\def\operator@font{\sf}\makeatother

\DeclareMathOperator*{\svd}{svd}

\newcommand{\ubar}[1]{\text{\b{$#1$}}}
\usepackage{amsthm}\theoremstyle{definition} 
\newtheorem{defn}{Definition}[section]
\newcommand{\beq}{\begin{equation}}\newcommand{\eeq}{\end{equation}}
\renewcommand{\vec}[1]{\boldsymbol{#1}}

\usepackage{lastpage,fancyhdr}
\usepackage{ifthen,hyperref}
\usepackage{lipsum}
\fancyheadoffset[L]{0.25in}
\fancyfootoffset[L]{0.25in}
\title{Singular knee identification to support emergence recognition\\in physical swarm and cellular automata trajectories}
\author{Imraan A.~Faruque and Ishriak Ahmed
 \thanks{I.~A.~Faruque is with Oklahoma State University, Stillwater, OK 74078 USA (e-mail: i.faruque@okstate.edu).}
 \thanks{Ishriak Ahmed is with Oklahoma State University, Stillwater, OK 74078 USA (e-mail: ishriak.ahmed@okstate.edu). }
\thanks{This work was supported in part by ONR Young Investigator Award N00014-19-1-2216. Copyright by the authors June 2024; this work may be under consideration for publication and copyright may be transferred without notice, after which this version may no longer be accessible.}}
\begin{document}

%\twocolumn[
%  \begin{@twocolumnfalse}
    \maketitle
    \begin{abstract}
      After decades of attention, emergence continues to lack a centralized mathematical definition that leads to a rigorous emergence test applicable to physical flocks and swarms, particularly those containing both deterministic elements (eg, interactions) and stochastic perturbations like measurement noise. This study develops a heuristic test based on singular value curve analysis of data matrices containing deterministic and Gaussian noise signals.  The minimum detection criteria are identified, and statistical and matrix space analysis developed to determine upper and lower bounds. This study applies the analysis to representative examples by using recorded trajectories of mixed deterministic and stochastic trajectories for multi-agent, cellular automata, and biological video. Examples include Cucker Smale and Vicsek flocking, Gaussian noise and its integration, recorded observations of bird flocking, and 1D cellular automata. Ensemble simulations including measurement noise are performed to compute statistical variation and discussed relative to random matrix theory noise bounds. The results indicate singular knee analysis of recorded trajectories can detect gradated levels on a continuum of structure and noise.
      Across the eight singular value decay metrics considered, the angle subtended at the singular value knee emerges with the most potential for supporting cross-embodiment emergence detection, the size of noise bounds is used as an indication of required sample size, and the presence of a large fraction of singular values inside noise bounds as an indication of noise.    \end{abstract}
 % \end{@twocolumnfalse}
%]

%\maketitle
%\section*{Abstract}\vfill\eject
% \tableofcontents

\section{Introduction}
Fields such as robotics or the science of autonomy need emergent behavior detection tools, but emergent behavior lacks a consistent definition that yields rigorous tests. A dominance of constructive approaches is used, in which a coordinated behavior is often used as a design goal. Example behaviors include formation; combined cohesion, velocity alignment, and collision avoidance; or convergence/consensus in a variable (most commonly heading in multi-agent systems). Convergence proofs may be available for an individual application system, and do not necessarily generalize to other dynamics or other motions, particularly for behaviors not well approximated by consensus.

This paper develops a method for distinguishing ordered and uncoordinated multi-agent interactions from experimental data (i.e., measured trajectories) to support emergent behavior detection from experimental data.  The approach builds on a definition grounded in complexity theory, and develops an embodiment-agnostic analysis approach implemented through digital compression techniques on robotic and biological examples of physical swarming as well as cellular automata.

\section{Previous work \& background}

\subsection{Embodied work}
Embodied emergence quantification tends to either focus most strongly on either the graph-theoretic properties or on the trajectories (performance) of such networks. Prominent examples include the ``small-world" network property \citep{watts1998smallWorldNetworkDynamics}, which has been found in many examples (such as \textit{C. elegans} neural networks and actor social networks), or ``scale-free networks" \citep{barabasi2003scaleFreeNetworks}, which showed that many natural and engineered networks could be identified with this fractal structure.

Conversely, considerable engineered design or biological discovery work has focused on trajectory performance analysis, most often within the context of designing a network to achieve consensus \citep{bullo2022_lns}.

While these disciplines are valuable in their own right, the connections between trajectories and the underlying graph network remain less mature.  Nowhere is this more pronounced than when trajectory measurements are used to infer structural properties, which has seen limited progress and is most often implemented by creating a parallel simulation that exhibits similar properties. For example, \citet{schaub2015slowSwitchingAssemblies} used a set of trajectories for early attempts at identifying the underlying connecting properties by comparing waves of slow-switching assemblies in neural structures with embodied implementations, showing a link in spectral properties.

\paragraph{Previous emergence quantification} %\textit{Zhang et al 2023, Cheng et al 2014, Qu et al 2017.}
Heuristic metric singular value decomposition (SVD) entropy has been used to assess ecological network complexity \citep{strydom2021svdentropy}, and spectral properties of a synaptic connectivity weight matrix to label emergence of slow-switching assemblies in neuronal networks \citep{schaub2015slowSwitchingAssemblies}. Defining emergence by quantifying the ``element of surprise'' of an observer who is fully aware of the design of the system has often been proposed \citep{ronald_design_2008}. The element of surprise's dependence on individual observer capability and the time varying nature of the condition make this path somewhat subjective. Some efforts in language theoretic approaches have emphasized model construction from definitions to reduce subjectivity \citep{kubi_towardFormalization_2003}. Granger causality has been used as an emergence definition, labeling a macro variable as emergent from micro variables if and only if it is Granger caused by microstates and is also Granger autonomous from microstates, where Granger-autonomous is defined in \citet{seth_emergence_2010}. 
Using an information-theoretic framework to compute emergence in life has been proposed \citep{gershenson_emergence_2023}, and the specific computation procedure for the emerging parameter requires a more precise definition, an awareness or estimation of probability density functions (PDFs), and an embodiment-specific generalization to apply to trajectory data. Specific applications in algorithmic information theory have monitored a bit string encoding observational data for plurality of drops in the bit string's Kolmogorov structure function, an improvement in generality that comes at the price of computability \citep{bedard_algorithmic_2022}, particularly when limited to measured trajectories.

\paragraph{Emergence in computational algorithms}
A review of computationally-oriented emergence definitions and methods to compute them is available in \citet{kalantari_emergence_2020} which focuses computer science applications.  The study of cellular automata and random Boolean networks used high (explicitly defined) and low order structures to conclude that emergent behavior occurs in intermediate regions of randomness (order/disorder) of interactions, quantified via an algorithmic parameter $\lambda \in [0,1]$ for which embodiment specific ranges are labeled ``edge of chaos" \citep{gutowitz1995edgeOfChaos,mitchell1993edgeOfChaos}.

\paragraph{System identification}
A workable emergence test must not rely strongly on a candidate internal model structure, making it different from traditional system identification approaches like principal component analysis, proper orthogonal decomposition, dynamic mode decomposition, or physics-informed dynamic mode decomposition, where a model set of candidate model structures may be tested against measured trajectories \citep{Juang1994sysIdBook,Ljung1999sysIdBook,morelli2006aicraftSysid,tischler2006sysIdBook,schmidDMD2010}. Our goal in this study does not necessarily require building a model; such a model-based approach may in fact constrain the long-term goal of generalizing the approach to multiple model classes.
\subsection{Complexity science and definitions}
%\textit{Definitions: 
Advances in foundational work in defining emergence for general systems include \citet{Ryan2007} and in more detail in \citet{prokopenko2009information}.  Following this approach, this approach first defines a macrostate $M$ and a microstate $\mu$ which differ by a resolution $\mathcal{R}_{[\mu,M]}$ and scope $\mathcal{S}_{[\mu,M]}$, having the three  constraints $\mathcal{R}_M\leq \mathcal{R}_\mu$, $\mathcal{S}_M\geq \mathcal{S}_\mu$, and $(\mathcal{R}_M,\mathcal{S}_M)\neq(\mathcal{R}_u,\mathcal{S}_\mu)$. Then, a straightforward definition is constructed.
\begin{defn}[Emergent property] For a macrostate $M$ and a microstate $\mu$, each having resolution and scope constraints $\mathcal{R}_M\leq \mathcal{R}_\mu$, $\mathcal{S}_M\geq \mathcal{S}_\mu$, and $(\mathcal{R}_M,\mathcal{S}_M)\neq(\mathcal{R}_u,\mathcal{S}_\mu)$, a property is \textbf{\textit{emergent}} if and only if it is present in a macrostate $M$ and it is not present in the microstate $\mu$.\label{d:emergent}\end{defn} 
\citet{Ryan2007} suggests that Defn.~\ref{d:emergent} leads to significant fundamental outcomes: emergence requires spatial or temporal extent, interaction ``structure'' acts as a constraint between variables, a Gaussian distribution not satisfying emergence, and both component dependence and nonlinearity being prequisites for emergence.

This definition, arising first in abstract complexity theory, is progress in making rigorous a previously subjective or philosophically-dominated concept, and led to initial ideas for emergence testing: that of compressibility tests quantifying dimensionality reduction \citep{Licata2017informationLossPrinciple, strydom2021svdentropy}. These paths and related compositional systems theory \citep{rosas2024emergenceSoftwareView} suggest that structure and its constraints will be especially important for detecting emergence, for example, the dimensionality loss of flocking agents relative to the more diverse motions of uncoordinated agents.

\subsection{Contribution of this paper}
The relatively abstract progress in complexity theory has not taken root in engineering or applied robotics research.
The primary contributions of this paper include
\begin{itemize}
    \item applying a singular value curve analysis approach to multi-agent and cellular automata system trajectory (typically spatiotemporal records) analysis for order and disorder separation to support emergence prediction and detection
    \item applying complexity theory emergence concepts and theoretical noise bounds on singular value decay curves to identify a transition point across varying structure levels to that can serve as a heuristic differentiator to support identifying emergent and non-emergent systems
    \item introducing eight singular value curve metrics to analyze order and unstructured noise of an underlying dataset, particularly knee angle, relative singular value, and the fraction of singular values outside theoretical noise bounds    
    \item establishing representative macrostate behavioral contours by computing the singular value metrics and their statistical variation on examples of simulated and experimental data to analyze emergence criteria, including examples of deterministic and Gaussian noise signals, passerine bird flock video, and cellular automata.\end{itemize}

\section{Methods and Approach}
\subsection{Motivating ideas \& overall structure}
\begin{figure}\centering    \includegraphics[width=0.38\textwidth]{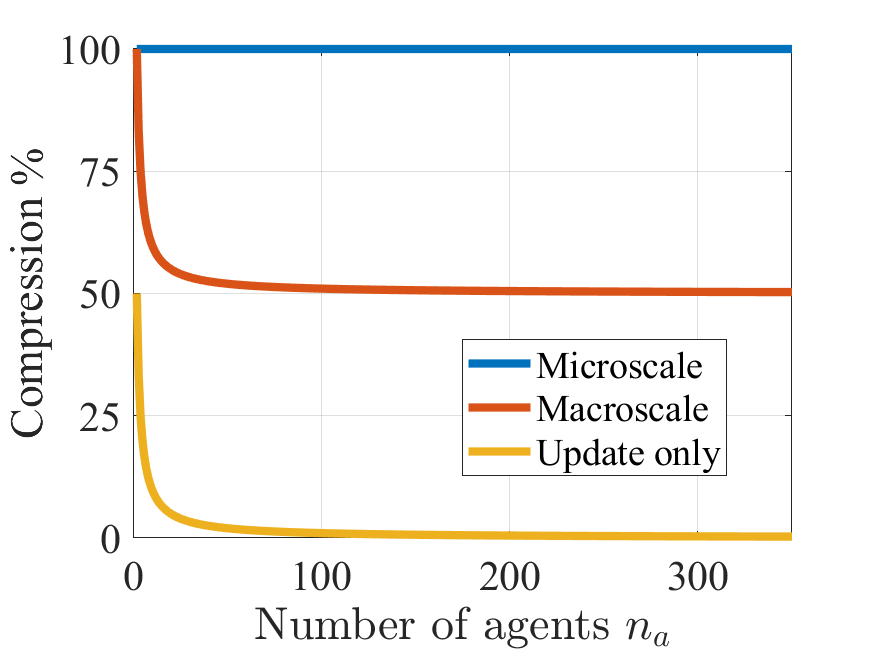}
    \caption{When $n_a$ distributed agents in 3D space (position and velocity) converge to a velocity-aligned flock, the macro scale can use the relative velocity bound to describe the behavior by simpler representations (fewer coordinates) than the $6 n_a$ needed for the microscale description.  For a swarm having reached velocity consensus in Euclidean space, one could specify the center of mass position (3 coordinates), converged velocity (3), and inter-agent vectors ($3 n_a$), or a total of $3n_a+6$ coordinates.  As $n_a$ grows large, the compression ratio approaches 50\%. By recognizing that converged inter-agent vectors are constant, updates could be to the center of mass position and velocity only, or 6 states. Macroscale compressibility then provides a stronger limit for large agent numbers.}\label{f:macroscaleCompressibility} \end{figure}

A motivating idea is to view imposed interaction rules between agents as constraining the agents to move within a subspace that provides a dimensionality reduction.  This reduction may be significant, as illustrated for the relatively strict case of perfect velocity alignment in Fig.~\ref{f:macroscaleCompressibility}.
A related idea is to recognize that the diversity of white noise provides a large number of individual components with predictable statistical characteristics (eg, bounds on singular values) and that passing such noise through a dynamic system (such as a coloring filter) results in added structure, a description that is more systematically explored through in this paper.

Relative to Defn.~\ref{d:emergent}, this study assumed resolution is fixed $R_M=R_\mu$ and address scoped $S_M>S_\mu$ by setting $S_M$ to cover the maximal spatial and temporal extent in a recorded trial. For a trial with $N_a$ agents measured in $d$ dimensional space at $N$ timesteps, the trial scope $S_M\propto N_a d n$. This study considers eight possible macrostate candidates, which are nonlinear computations of the $S_M$-scoped singular value decomposition curve.  These metrics are then discussed in the context of developing a heuristic emergence test based on comparing these macrostates to expected values.

An informed metric construction requires an expectation for these metrics and a measurement sufficiency condition measurement, for which this analysis considers the singular value decay curve and its behavior relative to noise and  numerical scope.  The Marcenko Pastur limits $(\bar{\sigma},\ubar{\sigma})$ are computed as a function of $\kappa$ to identify at which point the inflection.  E.g., this analysis begins to answer the question: when is a trial's measured scope $S_M$ sufficient to distinguish relative to noise?

\subsection{Dynamics}
For the physically embodied motions, this analysis first considered dynamics measured in the form \[\dot{\vec{x}} = \vec{f}(\vec{x},\vec{u}) + \vec{n},\] where multiple agents' states $\vec{x}$ varied according to a dynamic evolution rule $\vec{f}(\vec{x},\vec{u})$ quantifying both physics and inter-agent interactions as a function of input $\vec{u}$ and the signals are assumed to be measured with additive noise $\vec{n}$ of varying magnitude.  Specific $\vec{f}(\vec{x})$ definitions for eight example types are described in Section \ref{ss:simulationDynamics}. %Biological video recordings and cellular automata trajectories were similarly evaluated using this structure.

Emergence is often described in the context of cellular automata (CA) like Conway's Game of Life \citep{gardner1970conwayGameOfLife} and its extensions \citep{getz2022conwayLifeMultiState}. To test the analysis on cellular automata, the output trajectories from a 1D cellular automaton simulation having parametrically varying behavior were used as inputs. For a CA having a finite number of evolution rules, the parameter $\lambda \in [0,1]$, defined as the fraction of rules (not counting the all-dead rule) that lead to a living state, has been used as a parameter to vary behavior. Small $\lambda$ values give highly ordered patterns, while large values are chaotic \citep{gutowitz1995edgeOfChaos,mitchell1993edgeOfChaos}. Empirically, intermediate $\lambda$ values may reveal a region that includes interesting but not well-characterized behaviors commonly referred to as the ``edge of chaos.''

\subsection{Analysis approach}\label{s:skierAlgo}
In this approach, the two-dimensional positional time history is first arranged in a $2 n_a \times N$ matrix $X$, where $n_a$ is the number of agents in the swarm, and $N$ is the number of discrete timesteps the data is recorded. The matrix $X$ is de-biased by subtracting each row's mean and normalized by dividing each row of the centralized matrix with $\sqrt{N}\bar{s}_i$, where $\bar{s}_i$ is the standard deviation of the $i^{th}$ row to give the de-biased and normalized matrix $\tilde{X}$. The singular value curve is obtained by a singular value decomposition of $\tilde{X}$ and sorting the singular values in descending order. The knee position is determined with the triangle method and is used to determine heuristic metrics described in \ref{sec: SKIERmethods}.

\begin{figure}[htp]    \centering
    \includegraphics[width=0.4\textwidth]{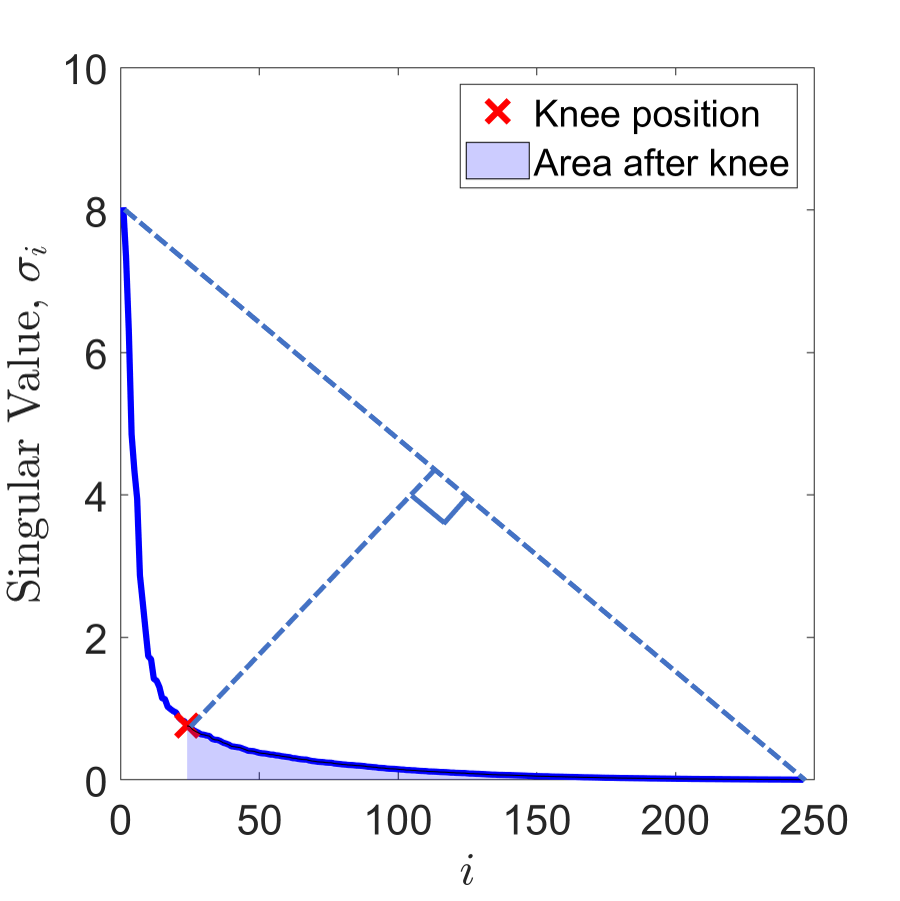}
    \caption{Singular values of a matrix are plotted in decreasing order to get the singular value curve. The values show a knee/elbow region. Generally after the knee region, the singular values do not decrease drastically as before. Heuristically, this point is used for `sufficient' low-rank reconstruction of a matrix.}\label{f:sv_SampleKnee}\end{figure}
    
\subsection{Singular value used for data interpretation}
Principal component analysis (PCA) and singular value decomposition (SVD) are two closely related techniques used for dimensionality reduction and feature extraction in linear algebra and machine learning. PCA can identify patterns in a dataset by transforming the data into a new coordinate system. The new coordinate frame aligns the first axis (the first principal component) with the direction of maximum data variance, the second axis (the second principal component) corresponds to the direction of maximum variance in the data that is orthogonal to the first axis, and so on.

The PCA of a data matrix $X$ can be implemented through SVD by subtracting from each row its respective row-wise mean to get $\tilde{X}$, computing the covariance matrix of the resulting de-biased data $\tilde{X}\tilde{X}^T$, and performing eigendecomposition on $\tilde{X}\tilde{X}^T$. The eigenvectors are the principal components, and the corresponding eigenvalues indicate the variance along each component. Alternatively, in
\beq USV^T = \svd(\tilde{X}),\eeq $U$ are the principal components and are also the eigenvectors of $\tilde{X}\tilde{X}^T$,
and the singular values taken in ascending order correspond to principal component order.

\subsubsection{Singular values \& their relationship to noisy structure}\label{s:whySingVals}
\paragraph{The singular value limit for Gaussian noise}
The singular values provided by an SVD of a data matrix are real numbers and are typically sorted from highest to lowest.

\begin{figure*}[ht]    \centering
    \begin{subfigure}[t]{0.32\textwidth}
    \includegraphics[width=\textwidth]{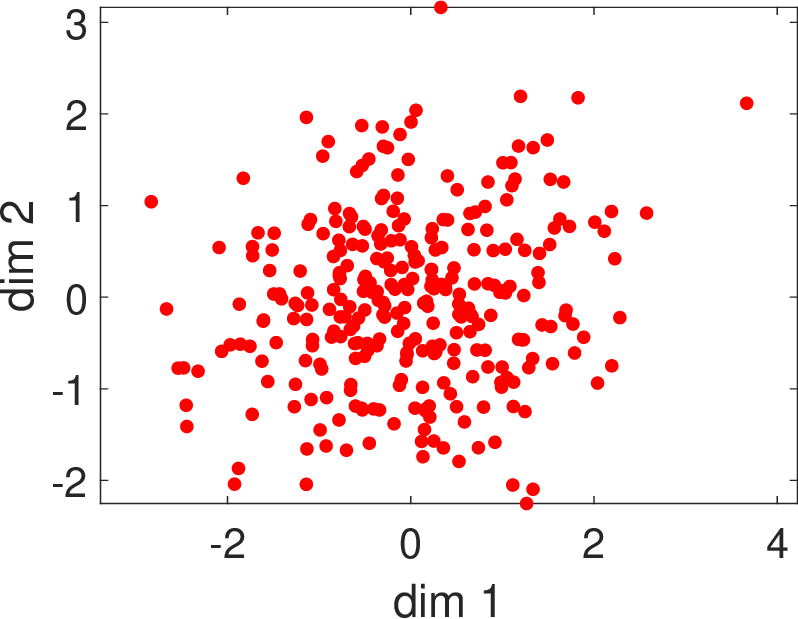}
    \caption{Points from $2\times300$ random matrix}\label{fig: noise2d}
    \end{subfigure}
    \begin{subfigure}[t]{0.32\textwidth} 
    \includegraphics[width=\textwidth]{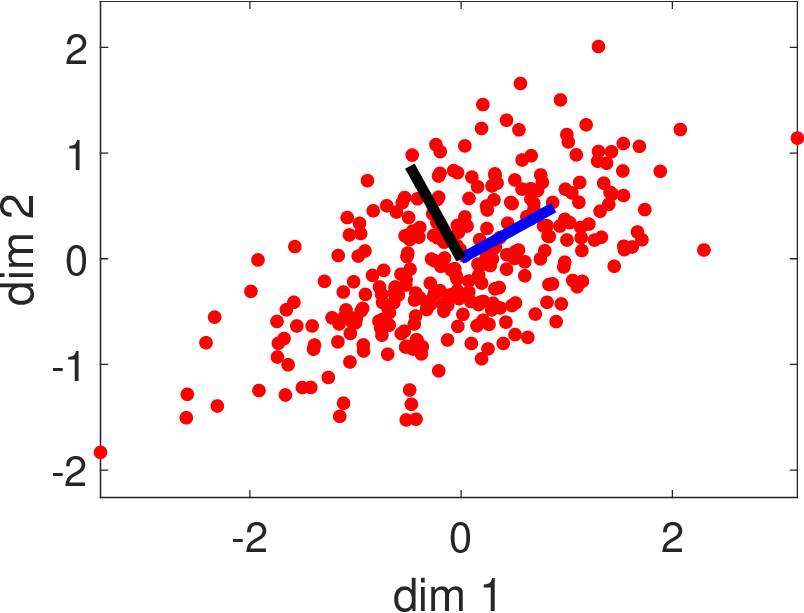}
    \caption{Points from $2\times300$ random matrix with correlation}\label{fig:noise 2d skewed}
    \end{subfigure}
    \begin{subfigure}[t]{0.32\textwidth} 
    \includegraphics[width=\textwidth]{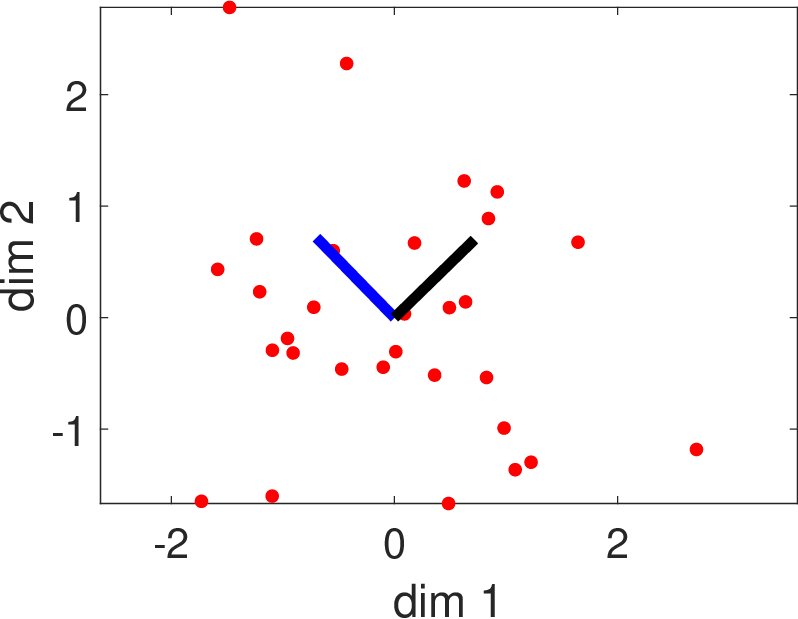}
    \caption{Points from $2\times30$ random matrix }\label{fig:noise2Dmisint}
    \end{subfigure}
    \caption{Singular values are used to determine the major axes in a PCA. In (b) the blue axis has a higher corresponding singular value than the black axis. In (a) there are no dominant singular values. In (c) though the data is random, due to the low sample size there are apparent principal axes.}
    \label{fig:comparisonPCASVD}
\end{figure*}
PCA analysis depends on the singular values to sort the components from most dominant to least dominant, suggesting that for a data matrix consisting of pure noise, the matrix's singular values are expected to be equal to each other. An illustration with two-dimensional data with 300 points is given in Fig. \ref{fig:comparisonPCASVD}. If the dataset does not provide enough points, one could misinterpret as seen in Fig.\ref{fig:noise2Dmisint} due to unequal singular values. 

The formal statement of this behavior is discussed in \citet{GavishOptSVDThreshold2014}. After proper scaling, the singular values of a random matrix all approach $1$.

Let $Y_n=\frac{1}{\sqrt{n}}\left(X_n+  Z_n\right )$, where $X_n$ is the data matrix and $Z_n$ is noise matrix where the matrix $Z_n$ has independent identically distributed zero mean entries. Each of the matrices has dimension $m\times n$, giving a row-to-column ratio $\kappa=m/n$. If X has rank $r$ one may write
\beq X_n=\sum_{i=1}^{r}x_i\vec{a}_{n,i}\vec{b}^T_{n,i},\eeq
where $a_{n,i}\in\mathcal{R}^m,b_{n,i}\in\mathcal{R}^n$, 
\beq Y_n=\sum_{i=1}^{m_n}y_{n,i}\vec{u}_{n,i}\vec{v}^T_{n,i},\eeq
and \begin{equation}\lim_{n\to\infty}y_{n,i}=\begin{cases}
  \sqrt{\left(x_i+\frac{1}{x_i}\right)\left(x_i+\frac{\kappa}{x_i}\right)} & \text{for }x_i>\kappa^{1/4}\\    
  1+\sqrt{\kappa} & \text{for }x_i\leq \kappa^{1/4}.    \end{cases}\label{eq:convTo1}\end{equation}
Here, if the data consists of only pure noise $Z_n$, $X_n=0$ which implies $x_i=0$, and as $\kappa \to 0$ the singular values should approach $1$. 

\paragraph{Bounds on the limit}
For a rectangular matrix $M\in\mathbb{R}^{m\times n}, (m<n)$ having entries drawn independently from a normal distribution with zero mean and unit variance, the limiting distribution of eigenvalues of the matrix $B=\frac{1}{n}MM^T$ is given by the Mar\u{c}enko-Pastur Law \citep{marcenko_distribution_1967}. When $n\to \infty$ with $\frac{m}{n}\to \kappa \in (0,1)$, the probability mass of eigenvalues of $B$ are bounded by $\kappa_{\pm}=(1\pm\sqrt(\kappa))^2$ \citep{bryson_marchenkopastur_2021}. Since the eigenvalues of $B$ are the singular values $\sigma_i$ of the matrix $\frac{1}{\sqrt n}M$, 
the singular values, $\sigma_i$ follows:
\[1-\sqrt\frac{m}{n}=\ubar\sigma^{rn} \leq \sigma_i \leq \bar\sigma^{rn}=1+\sqrt\frac{m}{n}.\]

This result provides a means to determine when an apparent mode/principal component of a data matrix differs significantly from random noise. Previous work has used this bound to denoise data by requiring dominant modes to have corresponding singular values greater than $\bar\sigma^{rn}$ \citep{veraart_denoising_2016}.  Singular values are thus intimately related to the inherent pattern of a data matrix.

\subsection{Analysis method and representative test cases}
This section develops a set of singular analysis curve metrics applicable to multiple domains in \ref{ss:dataAnalysis}.  These metrics are then applied to a set of multi-agent test cases across domains: physically embodied simulations (Section \ref{ss:simulationDynamics})with varying underlying dynamics scaling across random and structured motion including flocking and swarming models, biological flight records (\ref{ss:birds}), and cellular automata (\ref{ss:cellularAutomata}).

\subsubsection{Singular curve analysis}\label{ss:dataAnalysis}
This section describes singular knee identification (SKI) and subsequent analysis for order recognition from disorder to support emergence recognition (ORDER).  Consistent with Sec.~\ref{s:whySingVals}, each trajectory data matrix was first pre-processed by subtracting its row-wise mean and normalized by the individual row's variance.
\paragraph{Singular knee identification (SKI)}\label{sec: SKIERmethods} 
A singular value decomposition was performed on each of the trajectory data matrices, and the singular values $\sigma_i,~ i\in [1,2,... \mathrm{rank}(D)]$ were analyzed in decaying order (see Fig.~\ref{fig:SV_Gab}) to characterize the behavior of the singular value curve.

The ``knee'' position in the singular value decay curve was identified via a triangle method.  Given the decaying singular values $\sigma_i, i\in [1,2,... r_D], r_D=\mathrm{rank}(D)$, the triangle method connects the maximum singular value point $(1,\bar{\sigma})$ and minimum point $(r_D, \sigma_{r_D})$ with a line having slope $m_
\sigma=(\bar{\sigma}-\sigma_{r_D})/(1-r_D)$, and returns the datapoint $(i_k,\sigma_k)$ having maximum perpendicular distance to this line, i.e., the ``knee" of the curve.

\begin{figure}\centering    \includegraphics[width=0.42\textwidth]{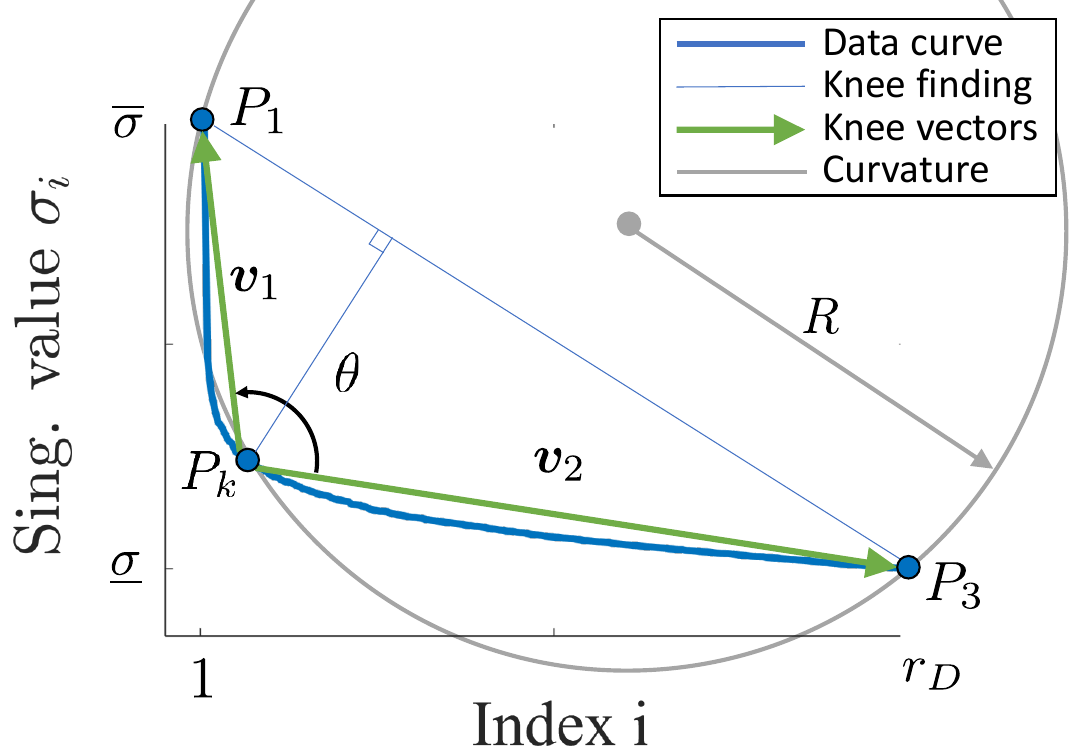}
    \caption{Singular value curve analysis: definitions of points $P_1, P_k, P_3,$ pre- and post-knee vectors $\vec{v}_1$ and $\vec{v}_2,$ and knee angle $\theta.$}    \label{f:svMetricDiag}\end{figure}

The following metrics were analyzed at the determined knee, based on the normalized singular value curve as shown in Fig.~\ref{f:svMetricDiag}.

\paragraph{Normalized singular value at knee}
Normalized singular value at knee is defined as $\sigma_k/\bar{\sigma}$, or the ratio of the singular value at knee $\sigma_{k}$ and the maximum (first) singular value $\bar{\sigma}=\max_i{\sigma_i}$.

\paragraph{Fraction of SVs outside random noise bounds}
This is determined by counting the number of singular values out of the range $(\ubar{\sigma}^{rn},\bar{\sigma}^{rn})$ and then dividing by the rank of the matrix (total number of singular values).

\paragraph{Knee location with respect to noise bounds}
This binary test returns 1 if $\sigma_{k}$ is outside $(\ubar{\sigma}^{rn},\bar{\sigma}^{rn})$ and 0 otherwise.

\paragraph{Normalized position of knee}
This is determined by dividing the knee index $i_{k}$ $\in (1, r_D)$ by $i_{max}=r_D$.

\paragraph{Normalized area beyond knee}
The normalized area both under the curve and beyond the knee is computed by first normalizing the singular values after knee as $y=\frac{1}{\bar{\sigma}}\begin{bmatrix}\sigma_{k}& \sigma_{k+1} &\cdots &\sigma_{r_D}\end{bmatrix}^T.$ The corresponding normalized post-knee indices
$x=\frac{1}{r_D}\begin{bmatrix}i_{k}& i_{k+1} &\cdots &i_{r_D}\end{bmatrix}^T$ allow finding the normalized area after knee by numerical integration (trapezoidal) of $\int y(x)dx$.

\paragraph{Knee angle}
The knee angle $\theta$ at knee is determined by the angle between the pre-knee vector $\vec{v}_1$ and post-knee vector $\vec{v}_2$ as \[\theta=\tan^{-1}\left(\frac{|[\vec{v}_1,0]\times[\vec{v}_2,0]|}{\left\langle\vec{v}_1, \vec{v}_2\right\rangle}\right),\] where $\times$ denotes the 3-dimensional vector cross product and $\langle\cdot{,}\cdot\rangle$ the inner product.

\paragraph{Curvature at knee}
The Menger curvature at the knee is determined by calculating the radius $R$ of the circle passing through the points $P_1=(0,1), P_k=(i_k/r_D,\sigma_k/\sigma_1)$, and $P_3=(1,0)$. 

More explicitly, curvature \[c=\frac{1}{R}=\frac{\sin(\theta)}{|-\vec{v}_1+\vec{v}_2|}.\]

\paragraph{Knee vector length ratio} The knee vector length ratio $V_\text{ratio}=\frac{|\vec{v}_2|}{|\vec{v}_1|}$ quantifies the length of the post-knee vector $\vec{v}_2$ relative to the pre-knee vector $\vec{v}_1$.

\subsubsection{Simulated Data Generation}\label{ss:simulationDynamics}
SKI was evaluated in two simulation classes to compare across differing dynamics and to provide statistical analysis. The number of agents $n_a$ and record length (number of discrete data points $N$) were chosen to match example bird experiments for efficient comparison. 
Each row of a simulation trajectory output contains a time history $x(t)$ or $ y(t)$ of a spatial dimension, thus the input data matrix is $2n_a\times N$.
For the systematic dynamic comparison, simulated datasets were the same time length (frames) and number of agents in the passerine flocking example.  For the statistical tests for different types of simulated data, five different initial conditions were run with five different simulation parameters to support computing the statistical variability; specifically the mean, median, standard deviation, and interquartile ranges.

\begin{figure*}    \centering \includegraphics[width=1\textwidth]{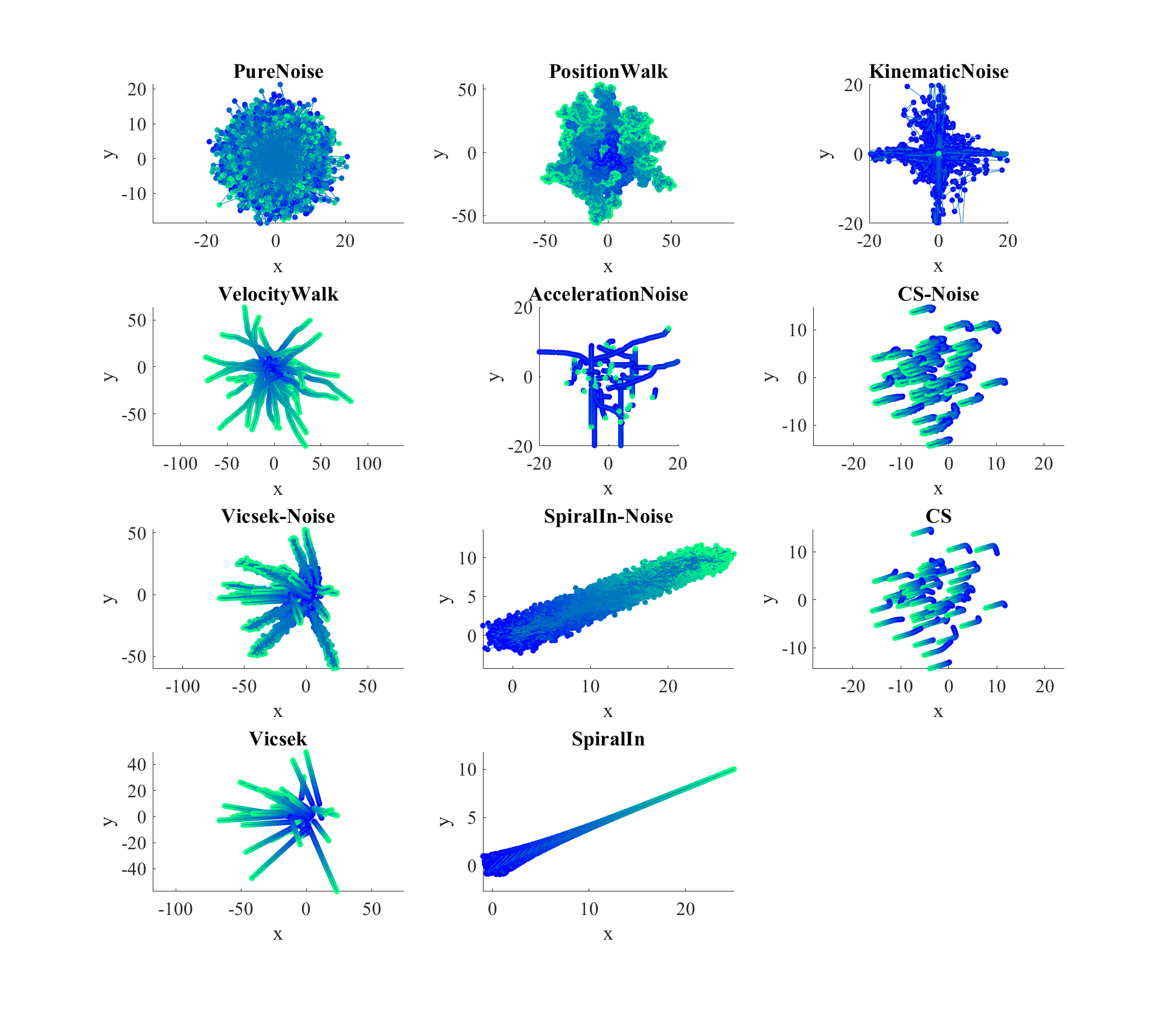}
    \caption{Example simulation trajectories (time progression is blue to green).} \label{fig:SimExamples}\end{figure*}

\paragraph{Pure noise}
The least structured example studied in this analysis is noise sampled from a normal distribution. Pure noise was modeled as a position update of agent $i$, in each timestep $t$, given by:
\begin{align}
    X_i(t+1)=\begin{bmatrix}x_i(t+1)\\y_i(t+1)\end{bmatrix}=n
\end{align} 
using samples $n\in \mathcal{R}^2$ from the normal distribution $\mathcal{N}(0,1)$.

\paragraph{Random walks}
One of the least structured motions is a random walk \citep{doyle1984randomWalks}, for which two cases were considered.  A ``position walk'' was defined in this study as
\begin{align} X_i(t+1)-X_i(t)&= n \label{e:positionWalk},\end{align} using samples $n$ from the normal distribution $\mathcal{N}(0,1)$, while a ``velocity walk'' was defined as
\begin{align}V_i(t+1)-V_i(t)&= n\\
X_i(t+1)-X_i(t)&=V_i(t)\Delta t.\end{align}

\paragraph{Kinematic and acceleration noise}
``Acceleration noise" denotes Gaussian noise inserted in acceleration as 
\begin{align}V_i(t+1)-V_i(t)&=\mu|V_i(t)| n\\
 X_i(t+1)-X_i(t)&=V_i(t)\Delta t, \label{e:accelNoiseDefn}\end{align} with constant $\mu\in [0,1]$ and again using noise samples $n$ from the random distribution $\mathcal{N}(0,1)$.  Acceleration noise is thus integrated twice to compute position, %, analogous to a coloring filter on turbulence providing forcing to a 
 while ``kinematic noise" was inserted only on velocity as 
 \begin{align} X_i(t+1)-X_i(t)&=\mu|X_i(t)| n\label{e:kinNoiseDefn}. \end{align}

\paragraph{Cucker-Smale model} 
\citet{cuckerSmale2007} model is a well-known model where the particles update their velocity based on inter-agent distance as follows:
$$V_i(t+1)-V_i(t)=\sum_{j=1}^{n_a} a_{ij}(V_j(t)-V_i(t)),$$
where, $$a_{ij} = \frac{K}{(1+||X_i-X_j||^2)^\beta}.$$ Here, $X$ represents agent positions and $K,\beta$ are simulation parameters.

\paragraph{Vicsek model}
The Vicsek model \citep{vicsek_novel_1995} holds that a particle $i$ traveling at a constant speed amongst $n_a$ agents updates its travel direction $\theta_i(t)$ as 
\[\theta_i(t+1) = \frac{1}{N}\sum_{\mathclap{\substack{i=1\\ j\neq i\\|X_i-X_j|\leq r}}}^{n_a} \theta_j(t),\]
responding to only the surrounding agents within a defined interaction radius $r$.
\paragraph{Spiral-In}
The inward spiral trajectories are defined by assuming agent $i$ at timestep $t$ follows
\begin{align}x_i(t)&=e^{-t}\sin{(2\pi (f+n_1)t+n_2)}+5t\\ 
y_i(t)&=e^{-t}\cos{(2\pi (f+n_1)t+n_3)}+2t,\end{align}
where $n_1,n_2,n_3$ represent Gaussian noise drawn from the standard normal distribution. This formulation corresponds to agent position being a sum of an inward spiral, an initial condition, and a convection towards $+x,+y$.

To corrupt a deterministic signal with simulated measurement noise, Gaussian noise with a standard deviation $5\%$ of the absolute difference between the temporal maximum and minimum was added.

\subsubsection{Biological data}\label{ss:birds}
To provide a more challenging experiment that better matches the level of observability a sailor or soldier might be placed in, we used a handheld camera phone video of passerine flocks outside a window on the Oklahoma State Campus. No camera calibration, framerate, motion model, background, image stabilization, or other motion parameters were provided to the algorithm; the video was the only input.  Background subtraction based on a moving window average was used to segment the targets, image subtraction (see \cite{ristroph2009automated,ristroph2010autostabilizer, saiful2021Visions,saiful2023SwarmDelayGroupSolo,ahmed2022hivistaBB,faruque2014wingMotionTransformation} for examples), and the data cropped to the longest period of simultaneous tracking.  Detected regions were indexed kinematically, and the raw measured values were provided to the algorithm. To place the results in context, the simulation cases were rerun for an equivalent number of agents and frames.  

\subsubsection{Cellular automata}\label{ss:cellularAutomata}

To collect a set of cellular automata trajectories, simulation result images were extracted from an online publicly accessible 1-dimensional CA, for which this study used David J Eck's javascript implementation \citep{Eck1DCA}, which provides a means to vary the simulator complexity parameter $\lambda$.  Each of the CA examples shares a single dead rule (``If a cell and all its neighbors are dead, then that cell is dead in the next world'') and includes an arbitrary number of pseudorandom rules.  Pseudorandom rules are generated via a Java-based random number generator \cite{BauSeedRandom} and a random assignment process (rule and environment function definitions are included in supplementary material).

To analyze cellular automata results, we used trajectories from the online simulator in \citep{Eck1DCA}, for which details are included in the Appendix. 
Simulations used a cell size of 5 pixels with 4 states, an isotropic neighborhood of size 5, and initialized all cells with a 50\% live/dead ratio.  An 8-bit grayscale color scheme represented 4 states and the simulation ran for 443 steps. 

The trajectory image was exported from the web applet, and the 1-pixel-wide vertical borders removed. The resulting image was subsampled by the cell size (5 pixels) in both directions giving a $443\times230$ image where each pixel represents an individual cell's state. This 2D array was transposed and became the input data matrix for singular knee analysis. 

\section{Results \& discussion}

\subsection{SV decay under noise}\label{ss:svDecayNoise}  
\begin{figure}[htbp]    \centering
\includegraphics[width=0.38\textwidth]{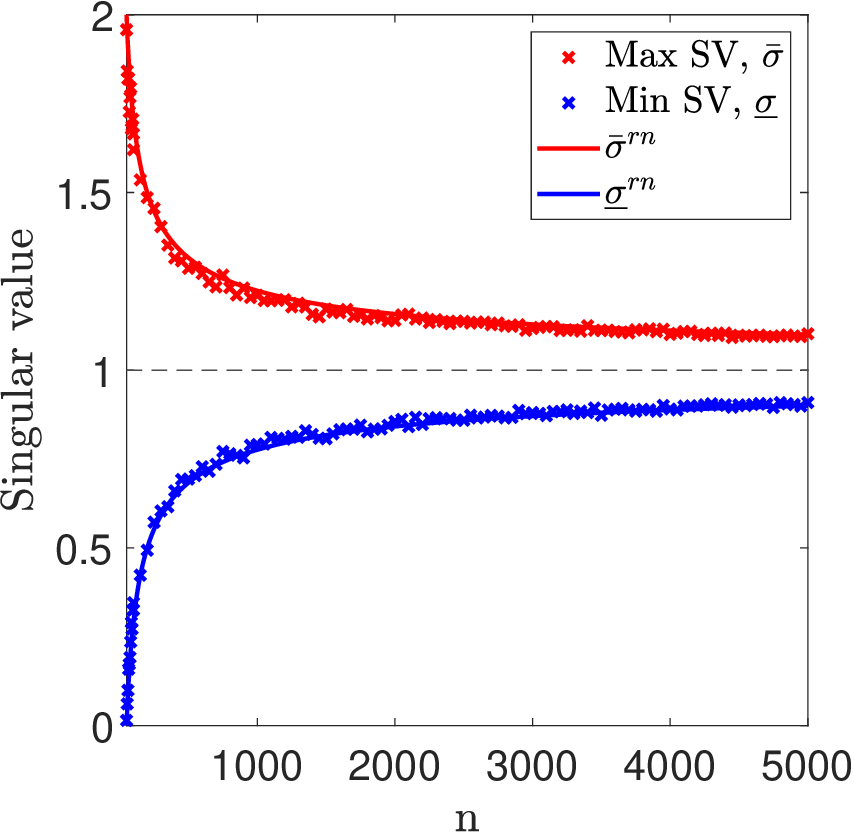}
    \caption{Singular value trend for simulated $50\times n$ random matrices defined by:  $\frac{1}{\sqrt{n}}$randn(50,n) for $n\geq50$. For low $n$, the maximum singular values are significantly higher than 1. The theoretical bounds can predict the bounds well.}   \label{fig:SV_trend}\end{figure}
    
We begin by numerically verifying the theoretical results discussed in Section \ref{s:whySingVals}.  We implement SVD on $50\times n$ matrices containing Gaussian noise for $n=1$ to $5000$ and illustrate the singular value decay curve in Fig.~\ref{fig:SV_trend}. At low $n$ the maximum and minimum singular values are well apart from 1, which we call the ``fluctuation bounds." \textit{In this region, noise deviations could be misinterpreted as patterns (principal components) in data.} For a given matrix, the singular values between the bounds signify the apparent principal components rising from a finite sample size which should have been equal to one.

\subsection{Ensemble simulation analysis}
\begin{figure*}[h!]
\centering
\begin{subfigure}[t]{0.3\textwidth}
\includegraphics[width=\textwidth]{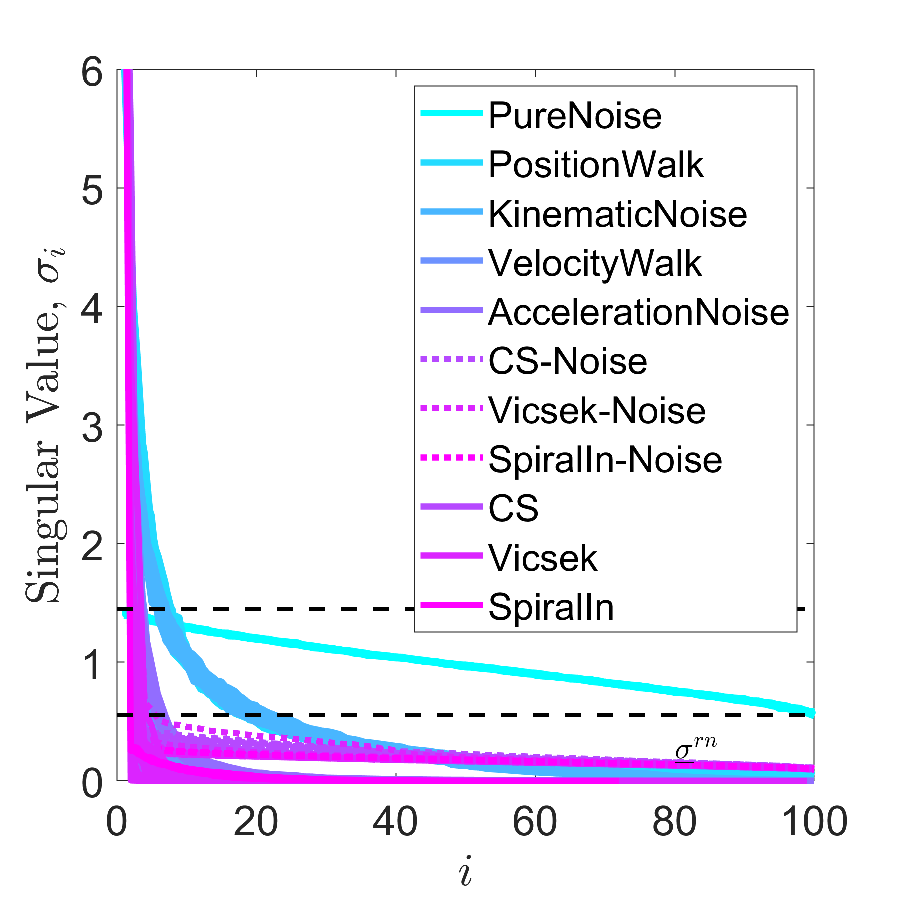}
\caption{Raw SVD}\label{f:svCurves225sims}
\end{subfigure}
\begin{subfigure}[t]{0.3\textwidth}
\includegraphics[width=\textwidth]{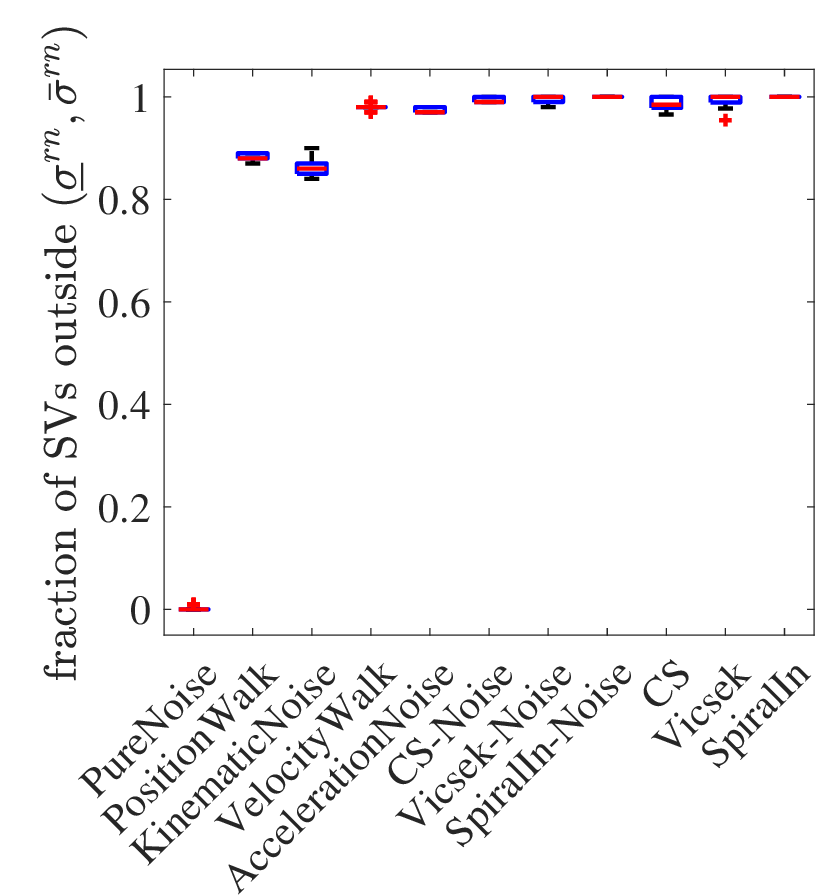}
\caption{SV's outside random noise bounds}\label{f:225sims_svs_outside}
\end{subfigure}
\begin{subfigure}[t]{0.3\textwidth}
\includegraphics[width=\textwidth]{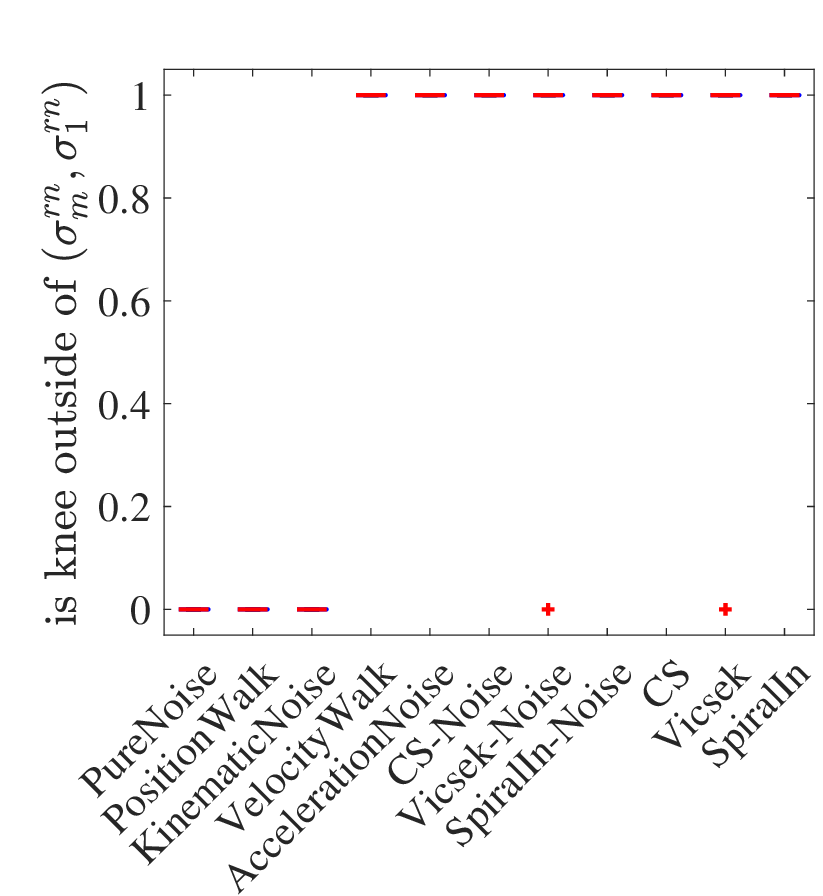}
\caption{Knee location with respect to noise bounds}\label{f:225sims_isKneeOutside}
\end{subfigure}
\begin{subfigure}[t]{0.3\textwidth}
\includegraphics[width=\textwidth]{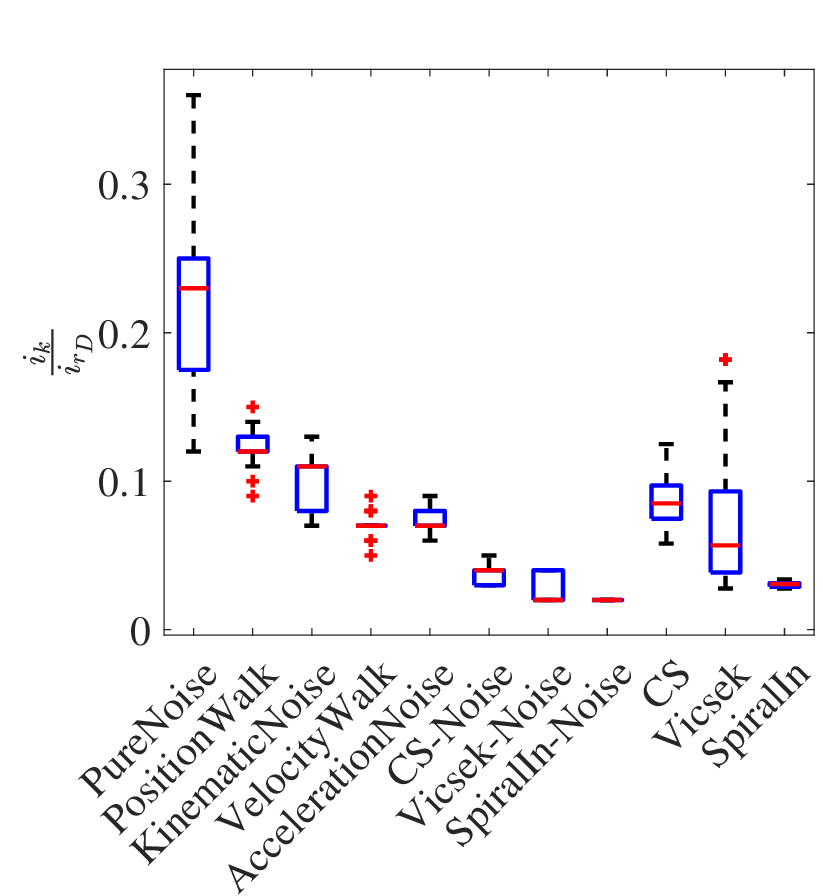}
\caption{Normalized position of knee}\label{f:225sims_norm_pos_knee}
\end{subfigure}
\begin{subfigure}[t]{0.3\textwidth}
\includegraphics[width=\textwidth]{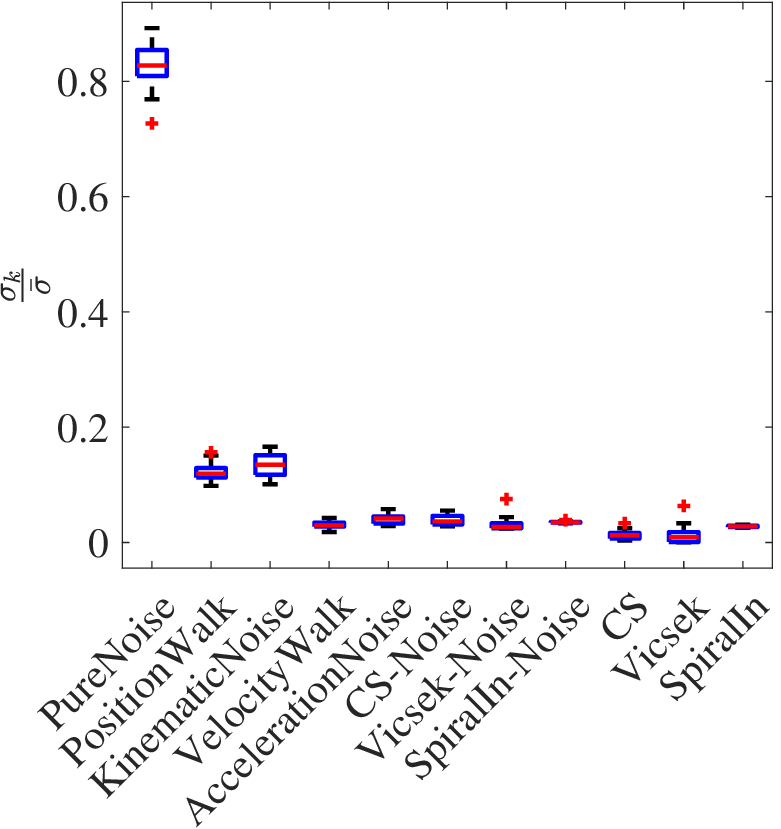}
\caption{Normalized singular value at knee}\label{f:225sims_norm_SV_knee}
\end{subfigure}
\begin{subfigure}[t]{0.3\textwidth}
\includegraphics[width=\textwidth]{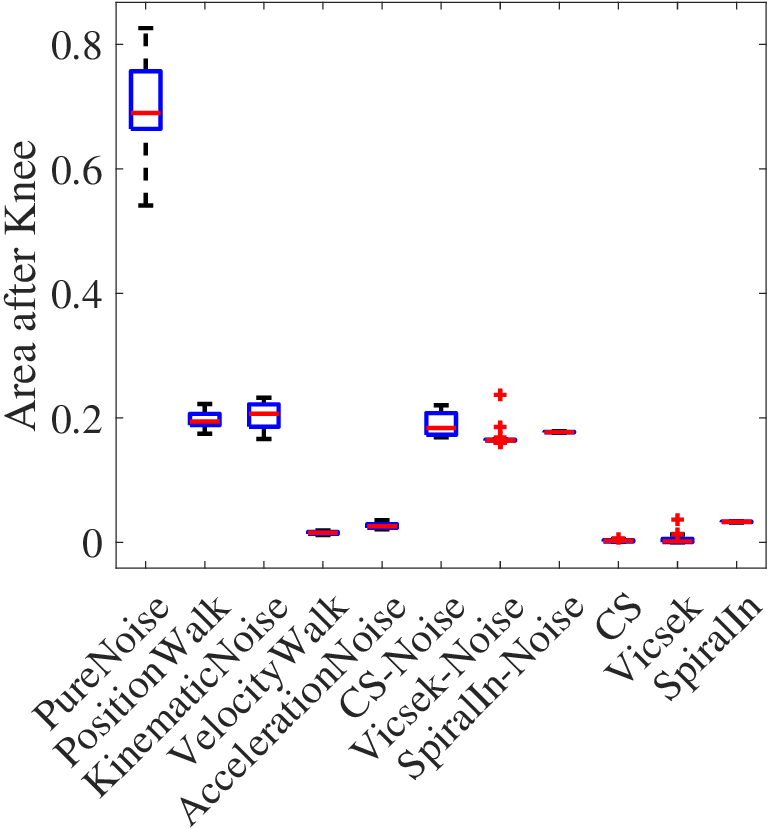}
\caption{Area after knee}\label{f:225sims_areaAfterKnee}
\end{subfigure}
\begin{subfigure}[t]{0.3\textwidth}
\includegraphics[width=\textwidth]{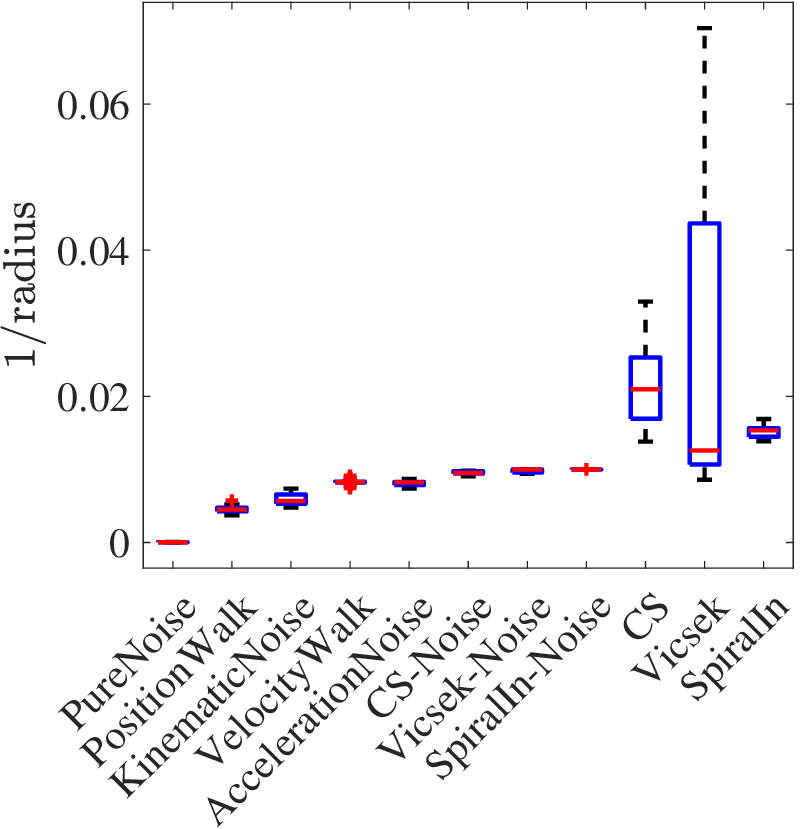}
\caption{Curvature}\label{f:225sims_curvature}
\end{subfigure}
\begin{subfigure}[t]{0.3\textwidth}
\includegraphics[width=\textwidth]{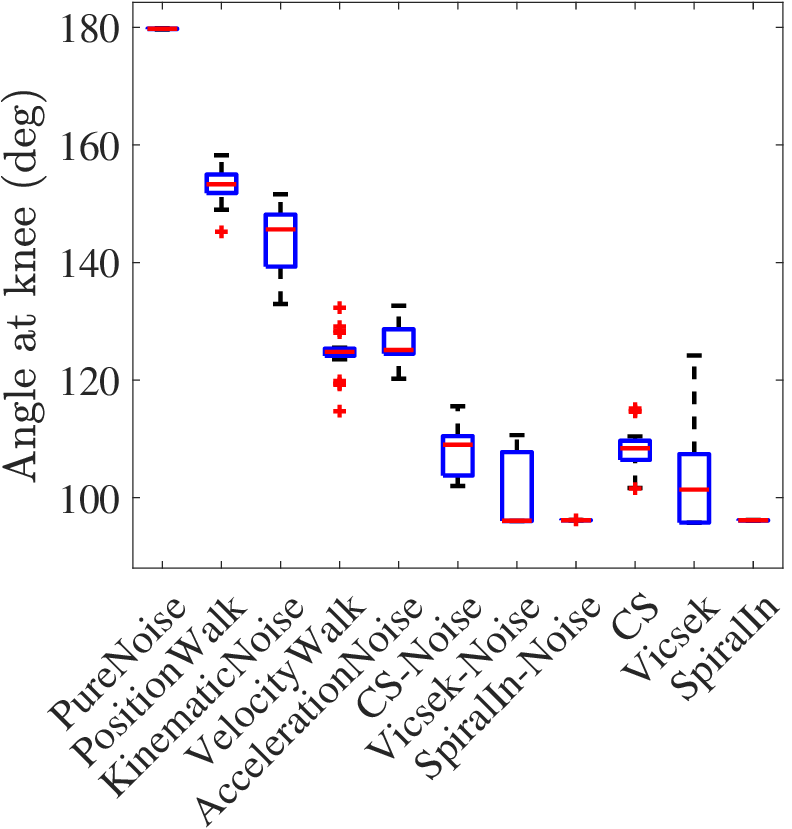}
\caption{Angle at knee}\label{f:225sims_AngleAtKnee}
\end{subfigure}
\begin{subfigure}[t]{0.3\textwidth}
\includegraphics[width=\textwidth]{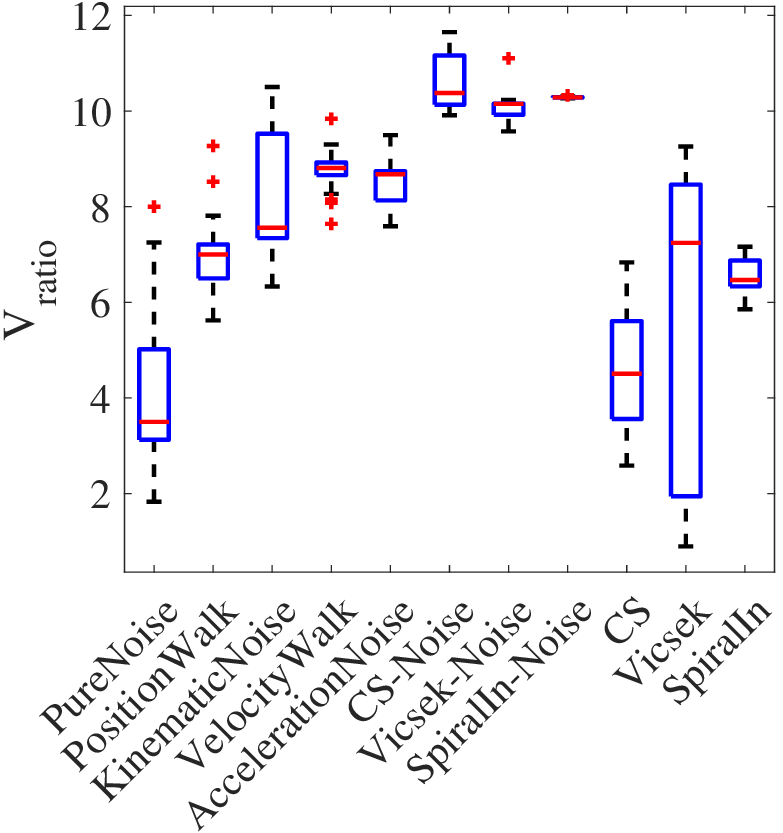}
\caption{Knee vector length ratio}\label{f:225sims_Vratio}
\end{subfigure}
\caption{Ensemble swarm \& flock motion simulation results, showing variation across 25 trials of each motion model. X-axis label ordering reflects a general trend from most disordered to most ordered.}
\label{fig: StatTestAll}\end{figure*}

To quantify the repeatability and robustness of the simulation results, this analysis considered 275 motion simulation instances (25 for each simulation model), with kinematic and acceleration noise magnitudes $\mu=0.3$. Each simulation has 50 agents with 500 timesteps. The singular value decay curves across 275 runs of these runs are shown in Figs.~\ref{fig: StatTestAll}. Kinematic and acceleration noise Eqn.~\eqref{e:kinNoiseDefn} may be rewritten to illustrate that kinematic and acceleration noise consists of multiplicative noise that decays with time, vs the additive noise seen in both kinds of random walks that provides a consistent stimulus and trajectories that do not necessarily decay.

\subsection{Bird flock tracking trajectories}
\begin{figure}[htb]\centering
    \includegraphics[width=0.38\textwidth]{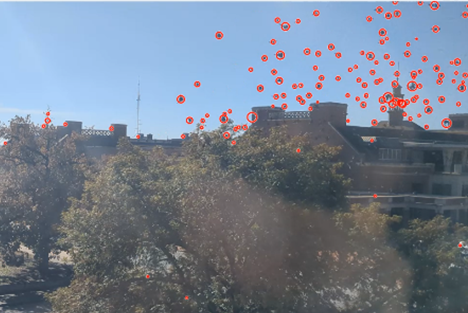}
    \caption{Example of visually-tracked flocking birds landing on an OKState campus tree.} \label{f:trackedBirdsGab}
\end{figure}

The visual algorithm was able to identify individual agents in the smartphone footage of flocking passerine birds, as seen in Fig.~\ref{f:trackedBirdsGab}. 
 The longest contiguous trajectory contained $n_a=18$ individual agents tracked across $N=41$ timesteps (frames).

\subsection{Cellular automata}
\begin{figure}[h!]\centering
\begin{subfigure}[t]{0.11\textwidth}
\includegraphics[width=\textwidth]{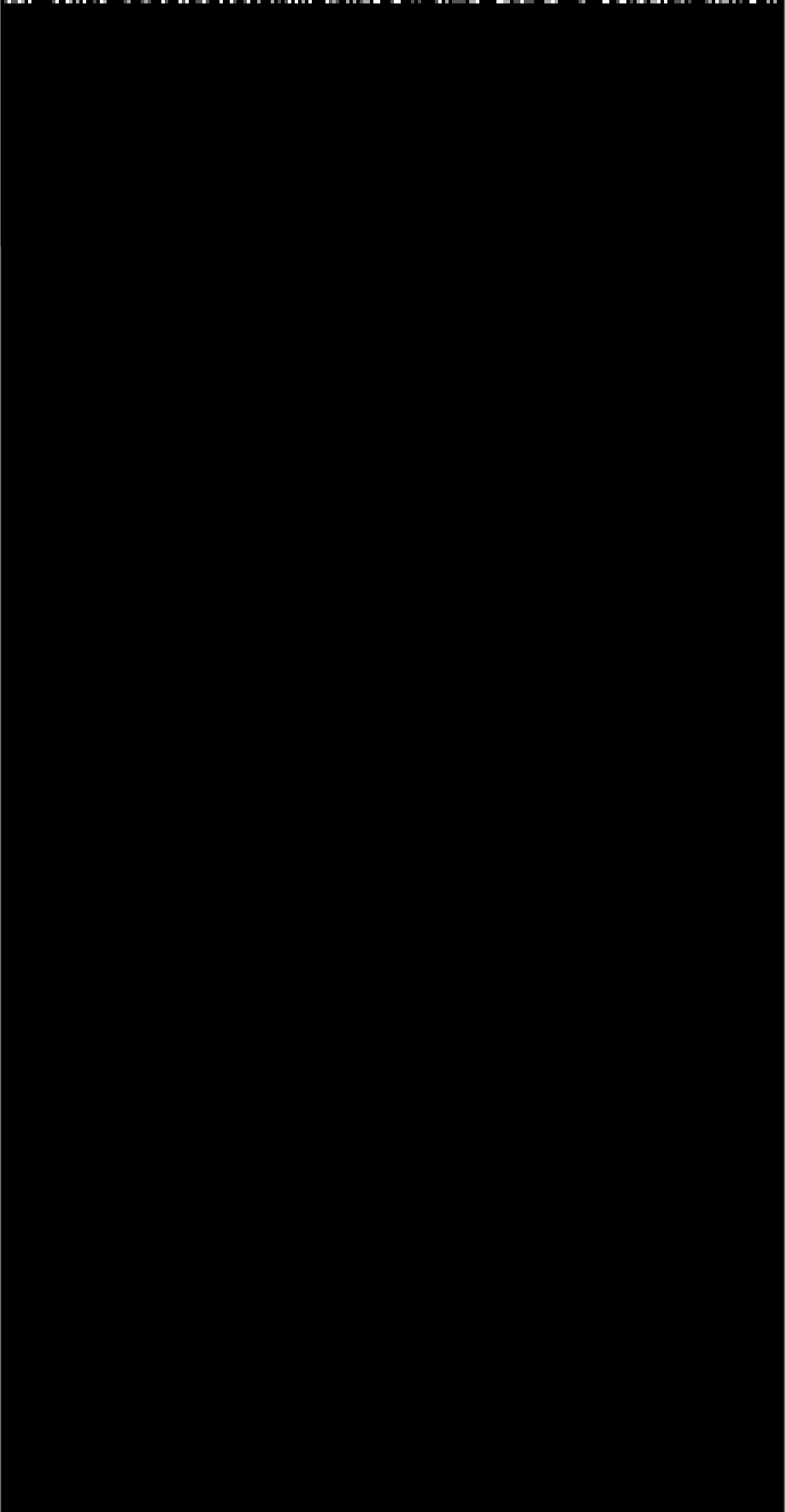}
\caption{$\lambda=0.00$}
\end{subfigure}
\begin{subfigure}[t]{0.11\textwidth}
\includegraphics[width=\textwidth]{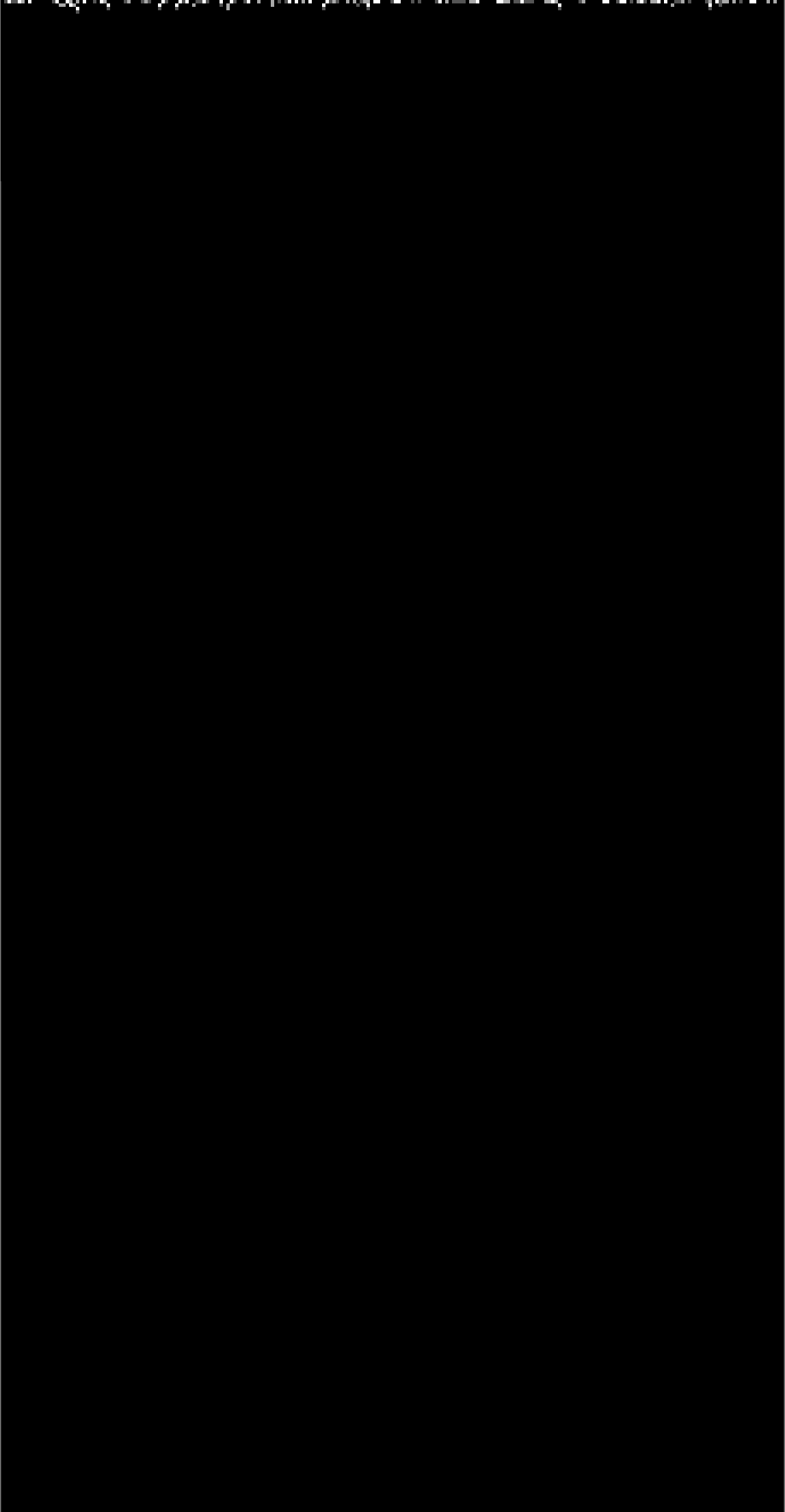}
\caption{$\lambda=0.10$}
\end{subfigure}
\begin{subfigure}[t]{0.11\textwidth}
\includegraphics[width=\textwidth]{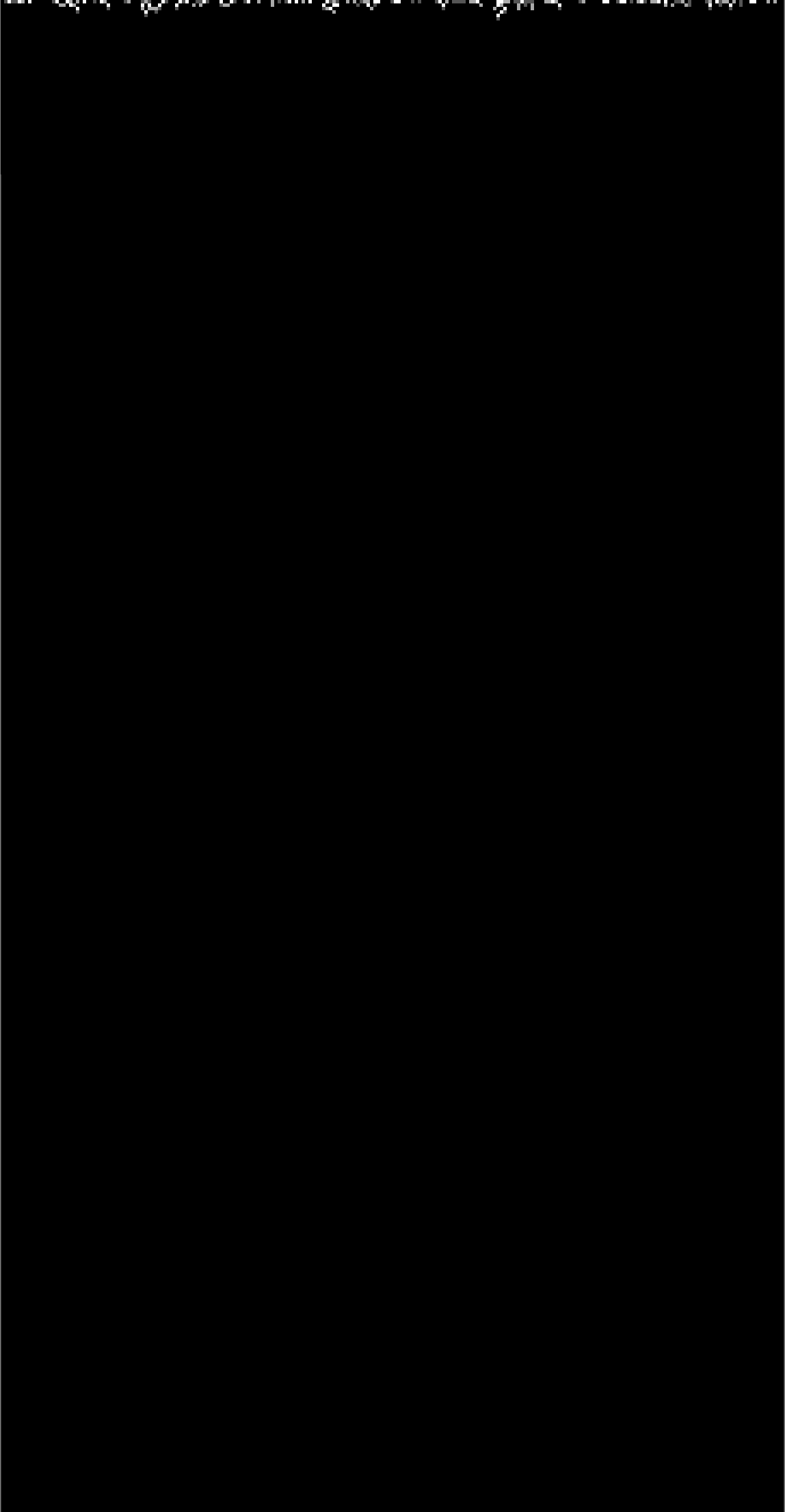}
\caption{$\lambda=0.20$}
\end{subfigure}
\begin{subfigure}[t]{0.11\textwidth}
\includegraphics[width=\textwidth]{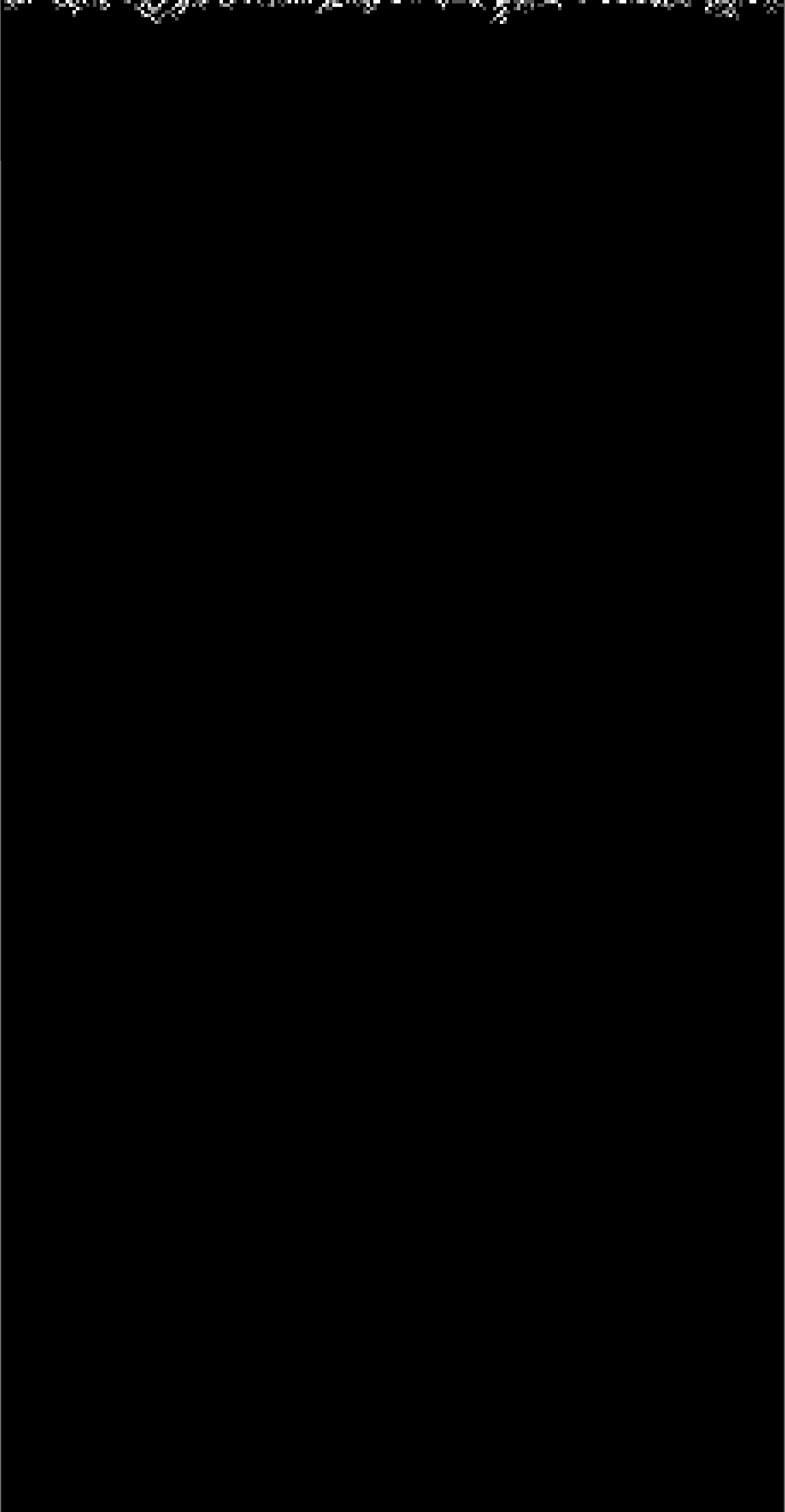}
\caption{$\lambda=0.30$}
\end{subfigure}
\begin{subfigure}[t]{0.11\textwidth}
\includegraphics[width=\textwidth]{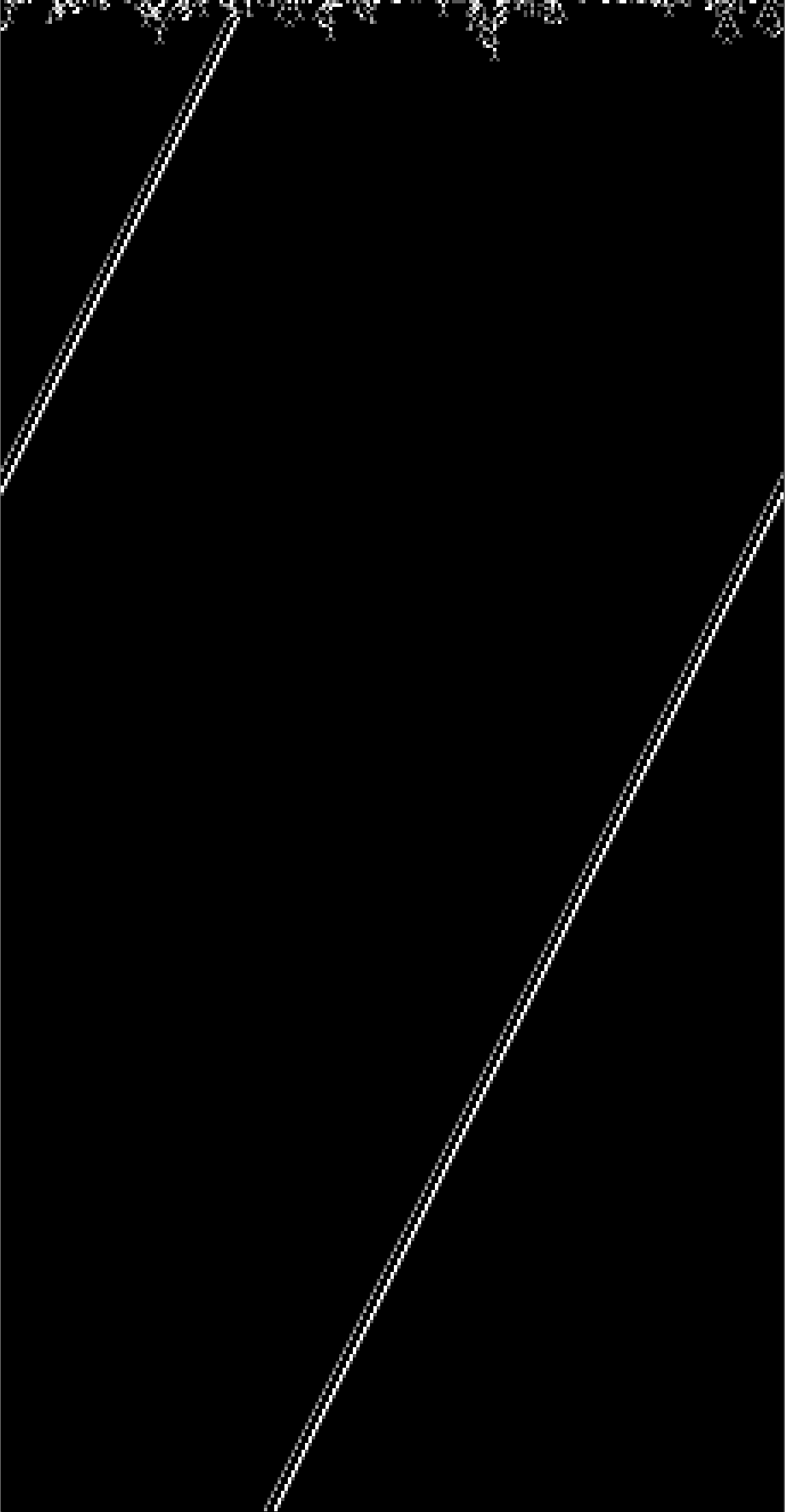}
\caption{$\lambda=0.37$}
\end{subfigure}
\begin{subfigure}[t]{0.11\textwidth}
\includegraphics[width=\textwidth]{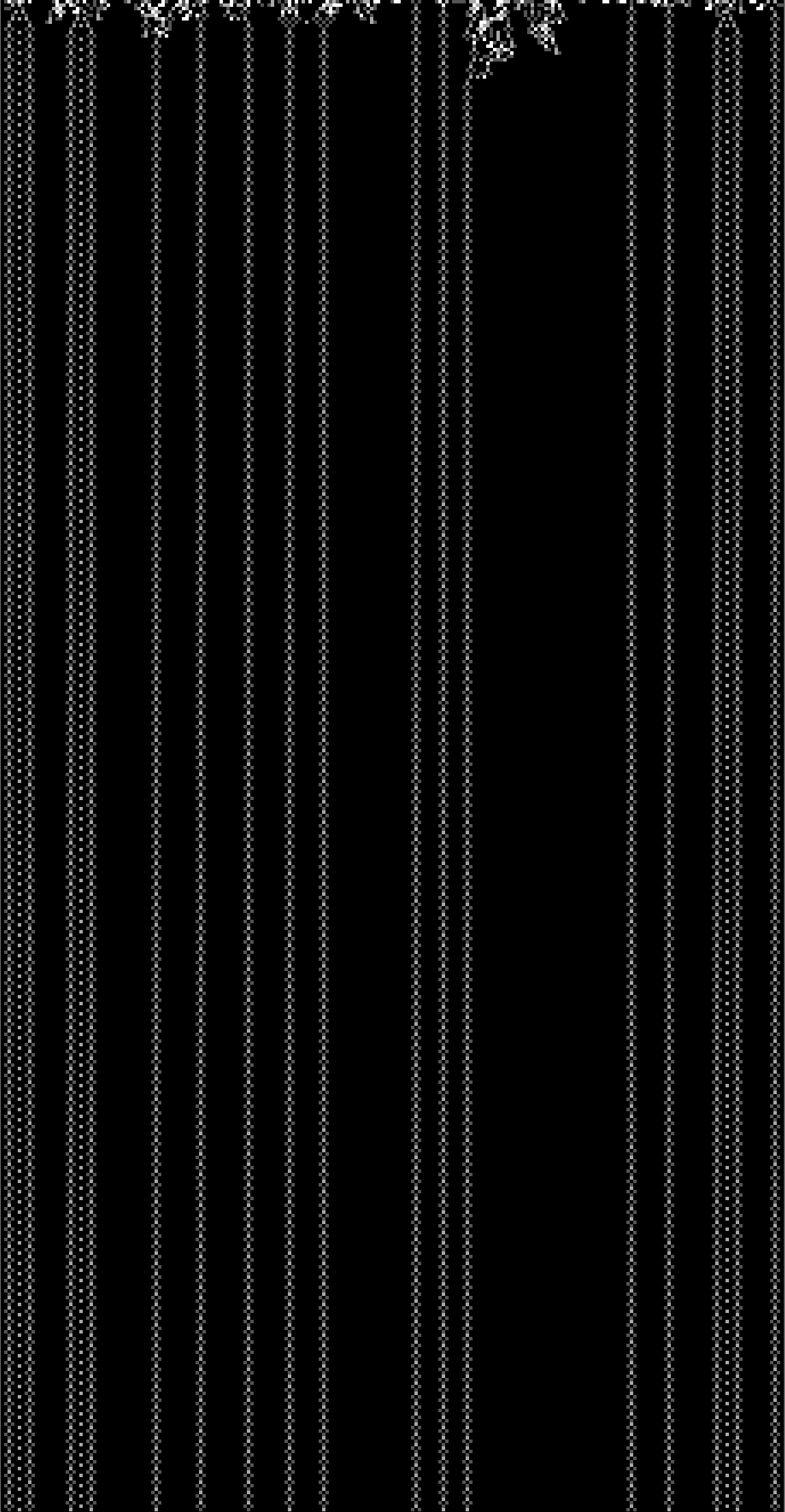}
\caption{$\lambda=0.40$}
\end{subfigure}
\begin{subfigure}[t]{0.11\textwidth}
\includegraphics[width=\textwidth]{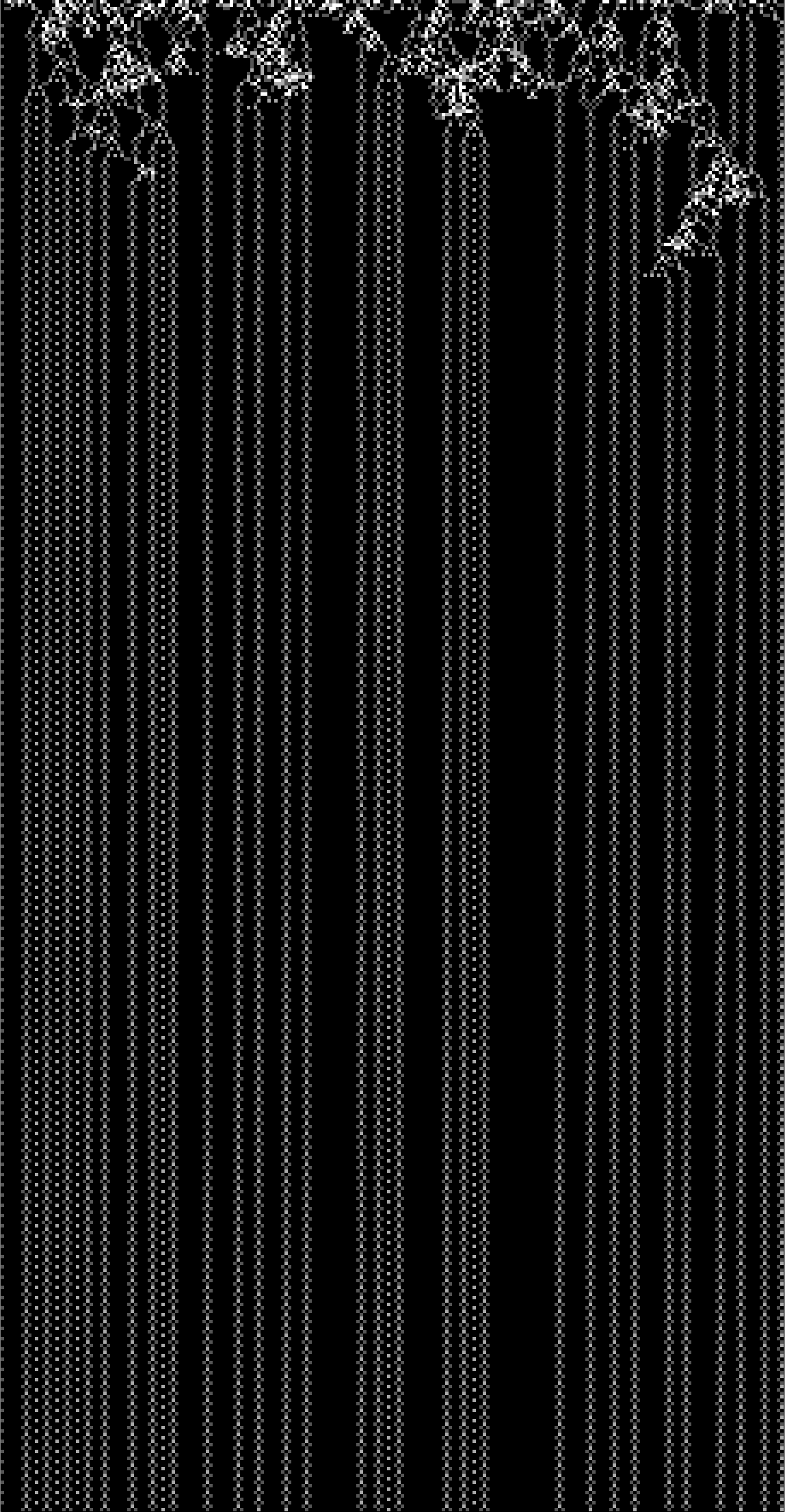}
\caption{$\lambda=0.50$}
\end{subfigure}
\begin{subfigure}[t]{0.11\textwidth}
\includegraphics[width=\textwidth]{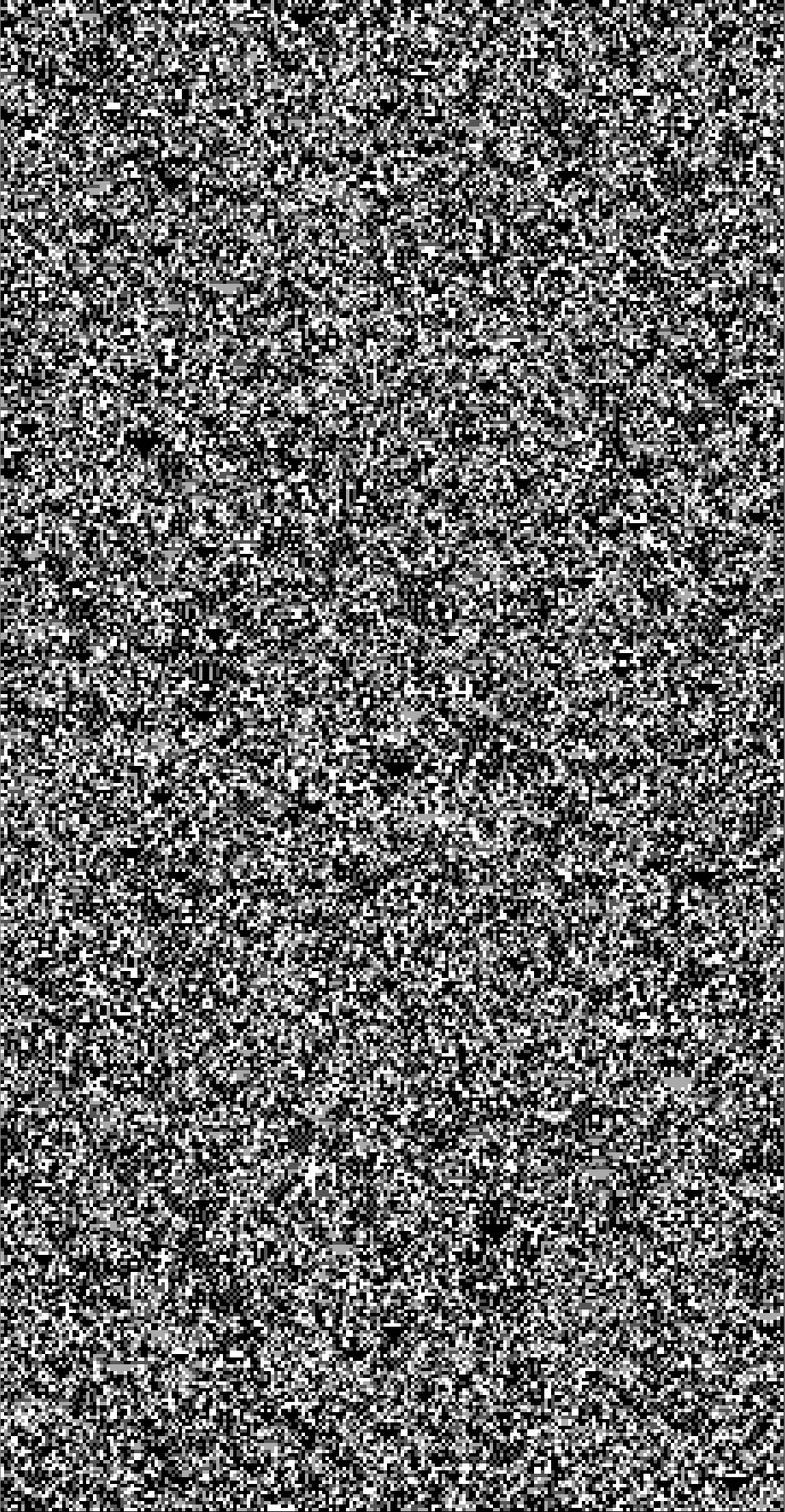}
\caption{$\lambda=0.65$}
\end{subfigure}
\begin{subfigure}[t]{0.11\textwidth}
\includegraphics[width=\textwidth]{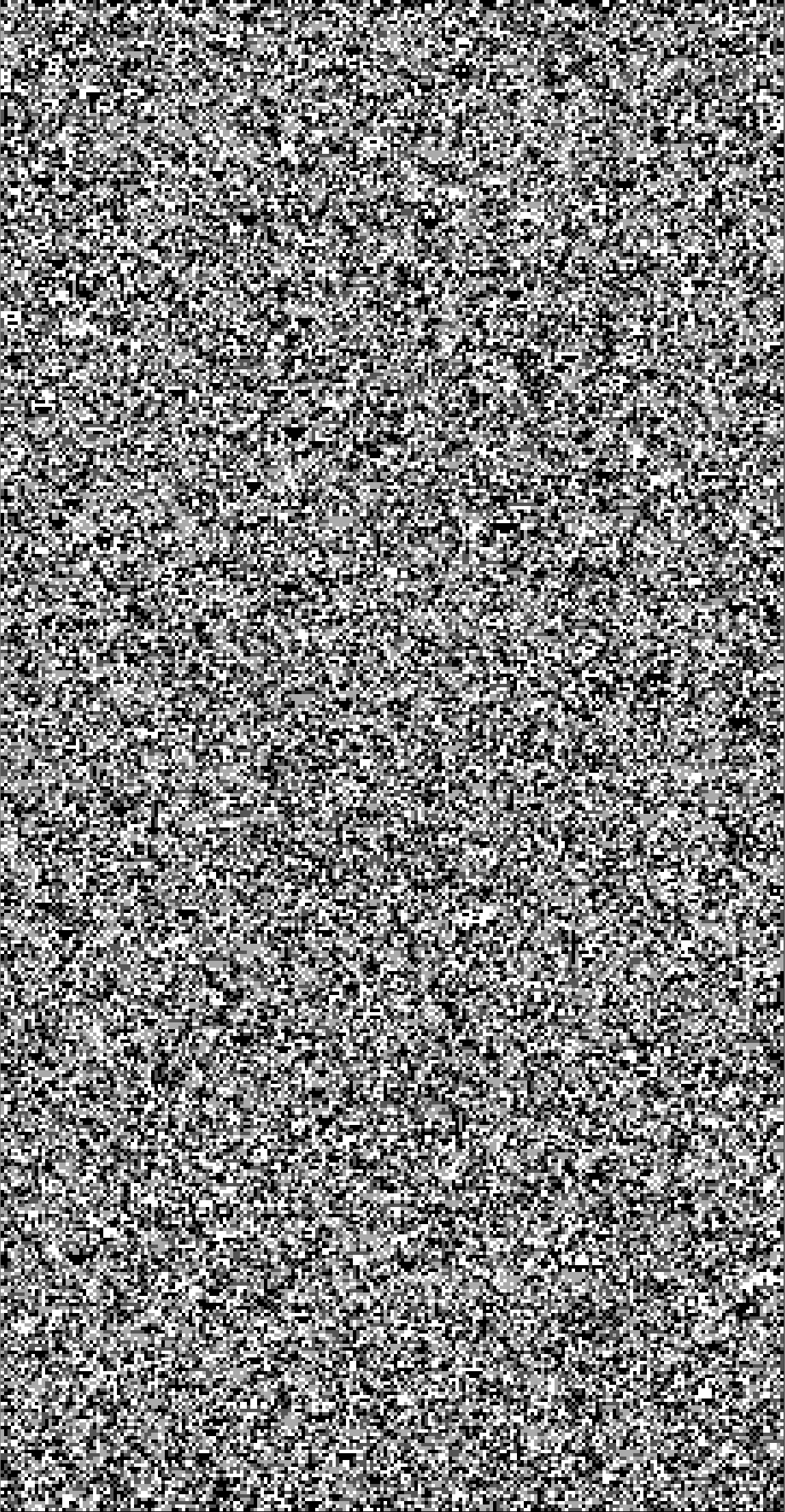}
\caption{$\lambda=0.75$}
\end{subfigure}
\begin{subfigure}[t]{0.11\textwidth}
\includegraphics[width=\textwidth]{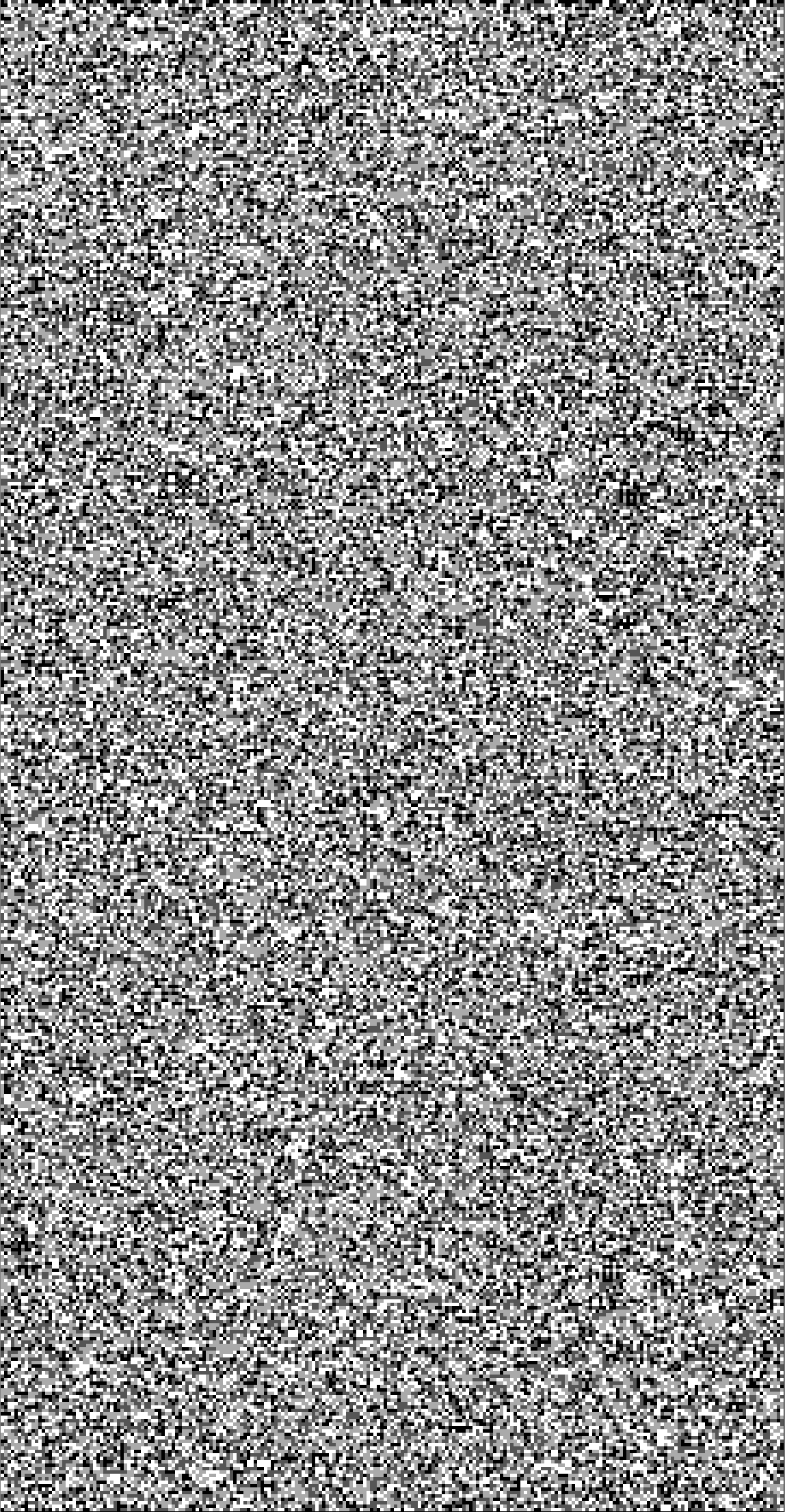}
\caption{$\lambda=0.85$}
\end{subfigure}
\begin{subfigure}[t]{0.11\textwidth}
\includegraphics[width=\textwidth]{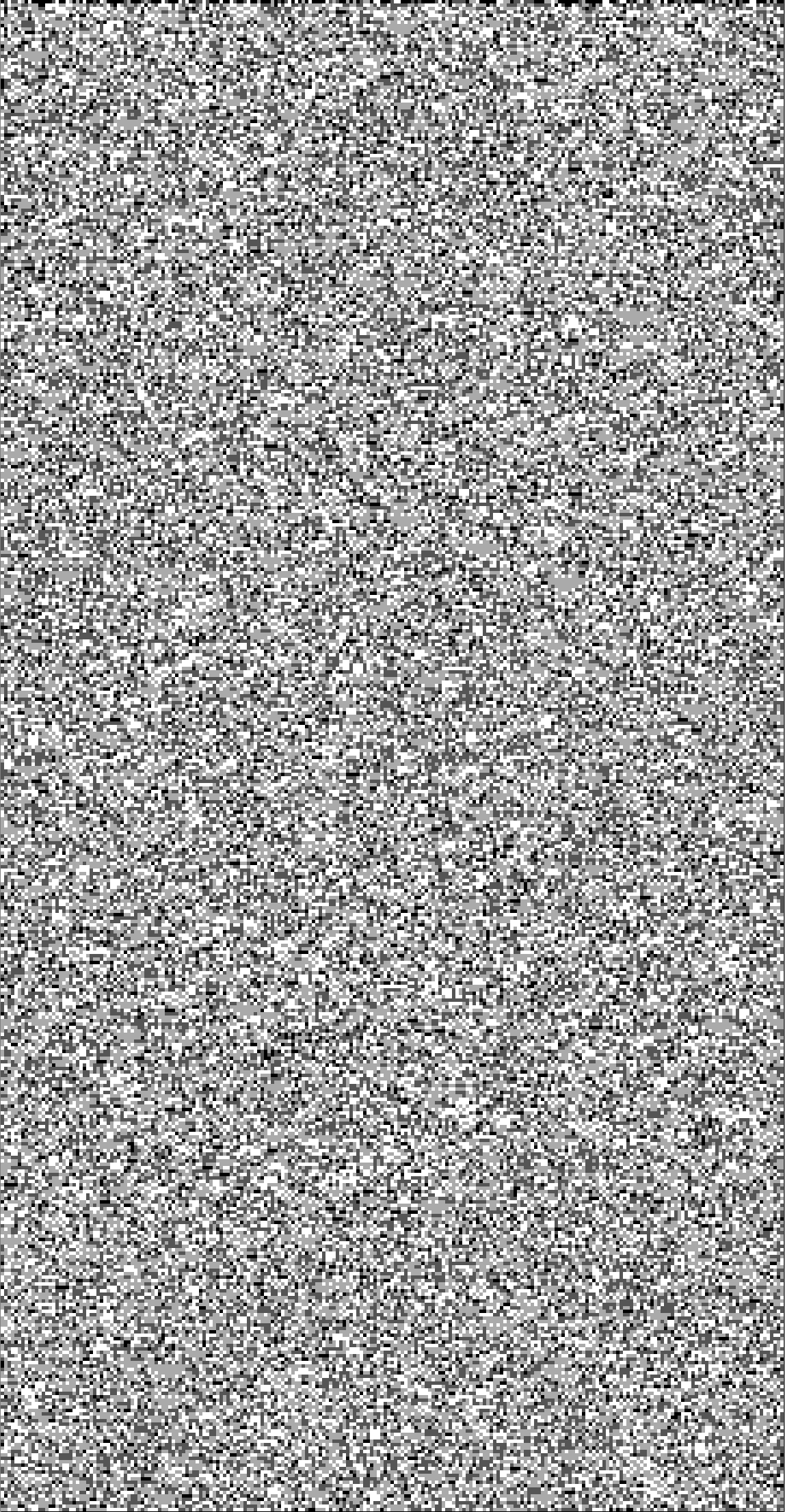}
\caption{$\lambda=0.95$}
\end{subfigure}
\begin{subfigure}[t]{0.11\textwidth}
\includegraphics[width=\textwidth]{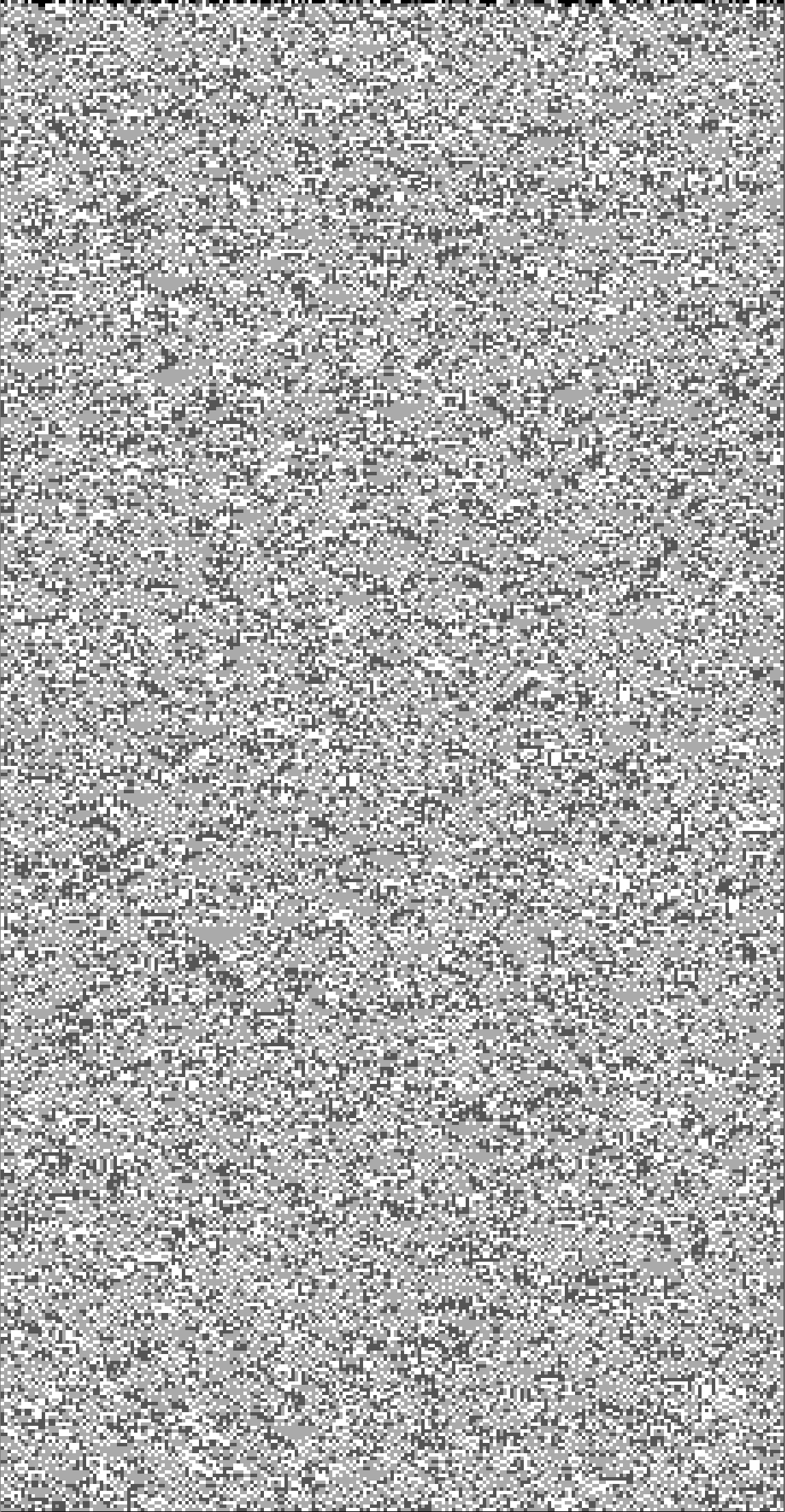}
\caption{$\lambda=0.99$}
\end{subfigure}

\caption{Cellular automata dataset examples}
\label{f: cellAutoData}
\end{figure}

Cellular automata were simulated for $\lambda$ = [0, 0.05, 0.1, 0.15, 0.2, 0.25, 0.3, 0.35, 0.37, 0.39, 0.4, 0.45, 0.5, 0.55, 0.6, 0.65, 0.7, 0.75, 0.8, 0.85, 0.9, 0.95, 0.99]. Regular patterns are visible in the resulting output for the 0.37-0.50 range in Fig.~\ref{f: cellAutoData}, a range subjectively identified as the ``edge-of-chaos'' \citep{mitchell1993edgeOfChaos,gutowitz1995edgeOfChaos}. 

\begin{figure*}[ht!]\centering
\begin{subfigure}[t]{0.3\textwidth}
\includegraphics[width=\textwidth]{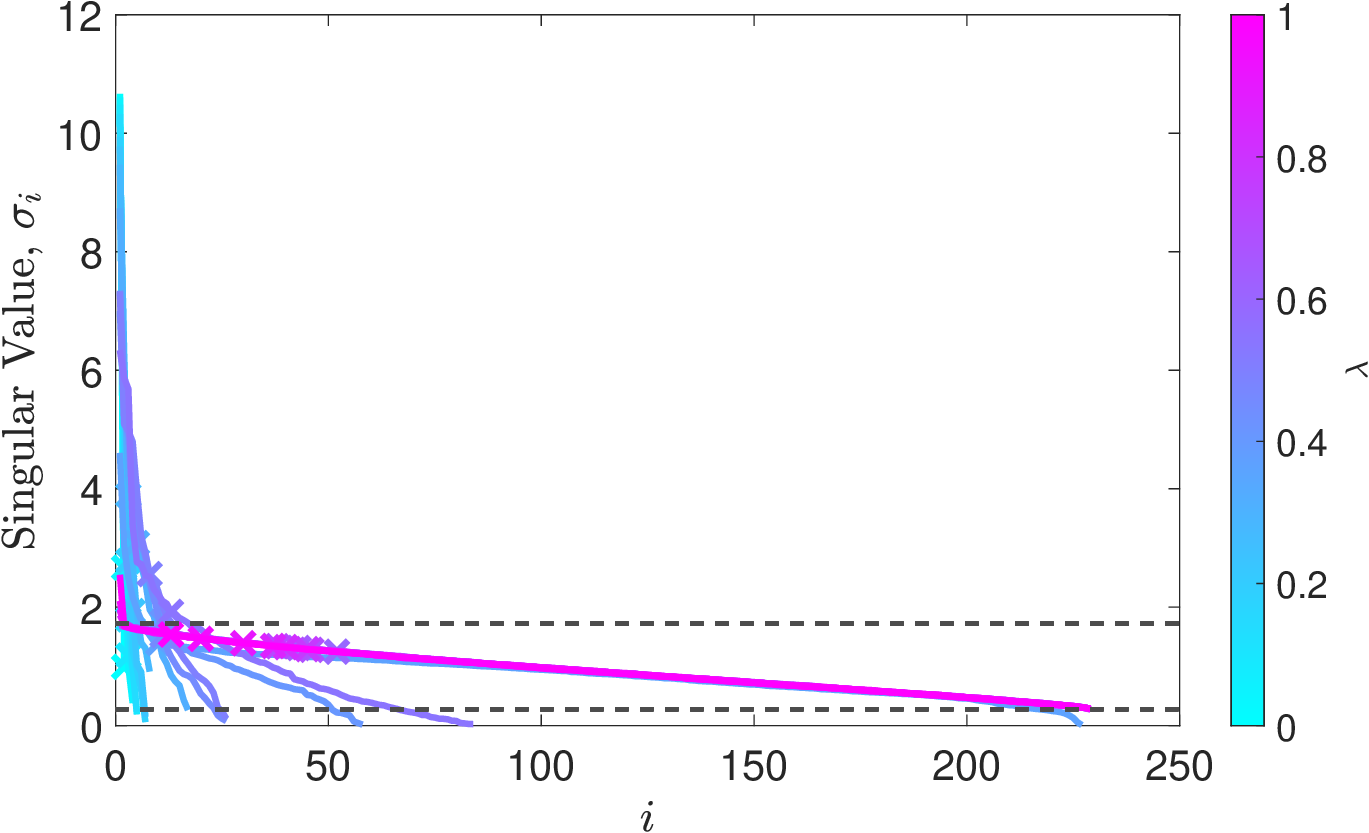}
\caption{Raw SVD}\label{f:cellAuto_svCurves}\end{subfigure}
\begin{subfigure}[t]{0.3\textwidth}
\includegraphics[width=\textwidth]{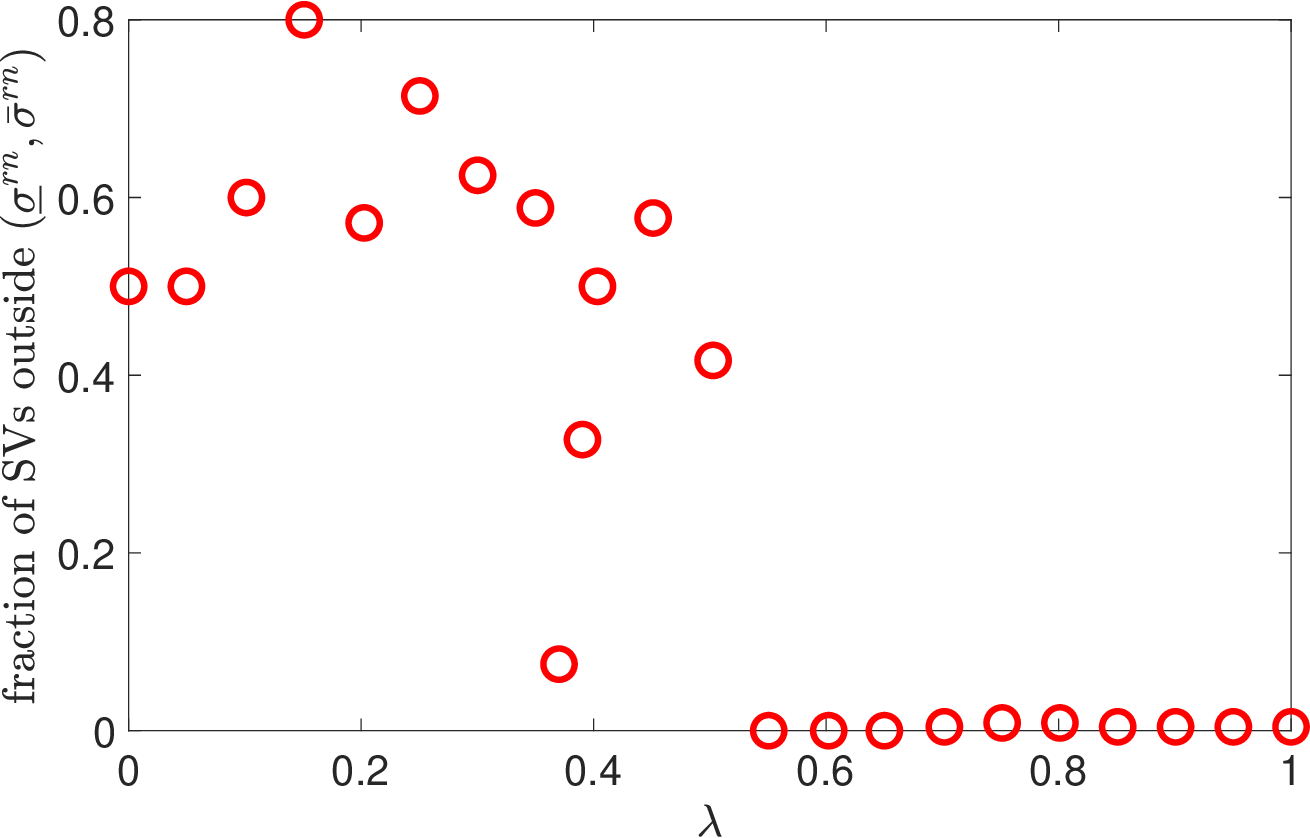}
\caption{SV's outside random noise bounds}\label{f:cellAuto_svs_outside}\end{subfigure}
\begin{subfigure}[t]{0.3\textwidth}
\includegraphics[width=\textwidth]{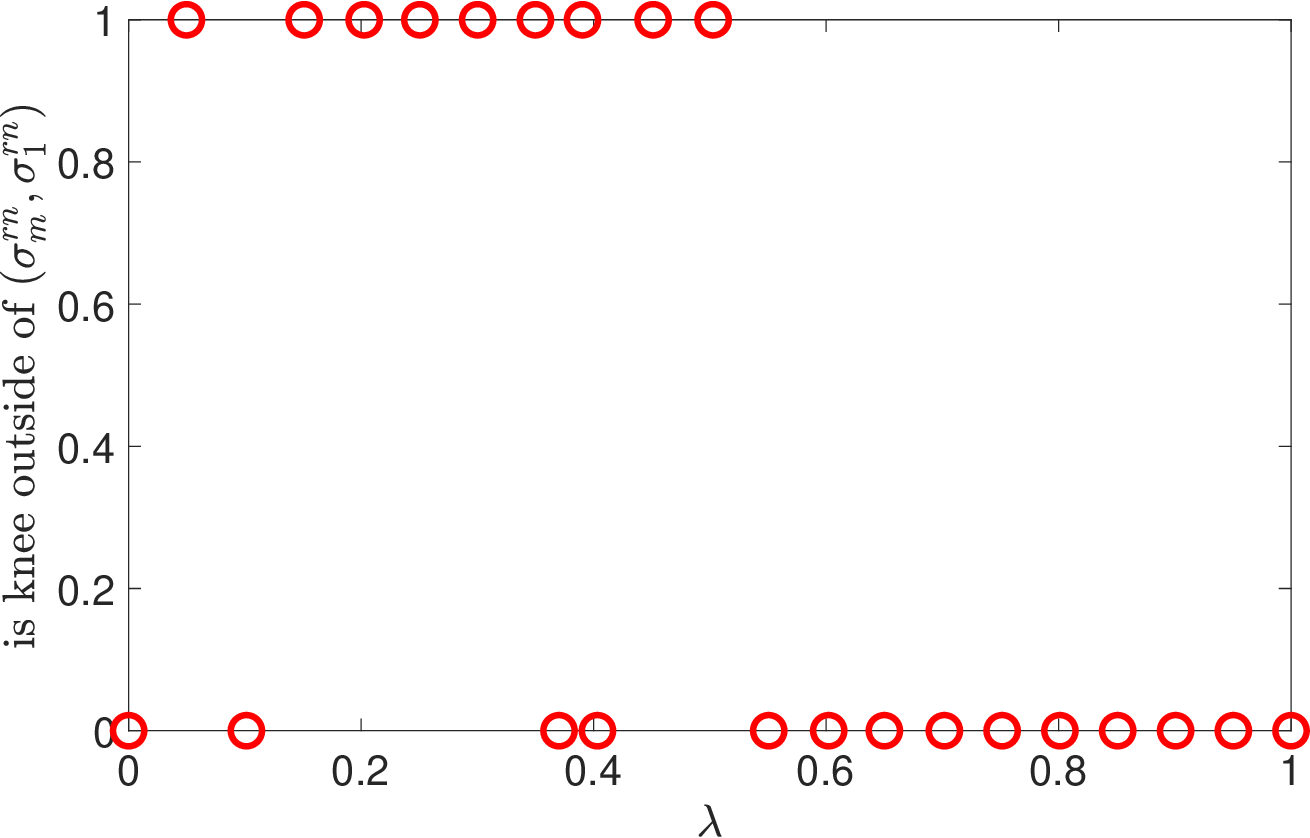}
\caption{Knee location with respect to noise bounds}\label{f:cellAuto_isKneeOutside}\end{subfigure}
\begin{subfigure}[t]{0.3\textwidth}
\includegraphics[width=\textwidth]{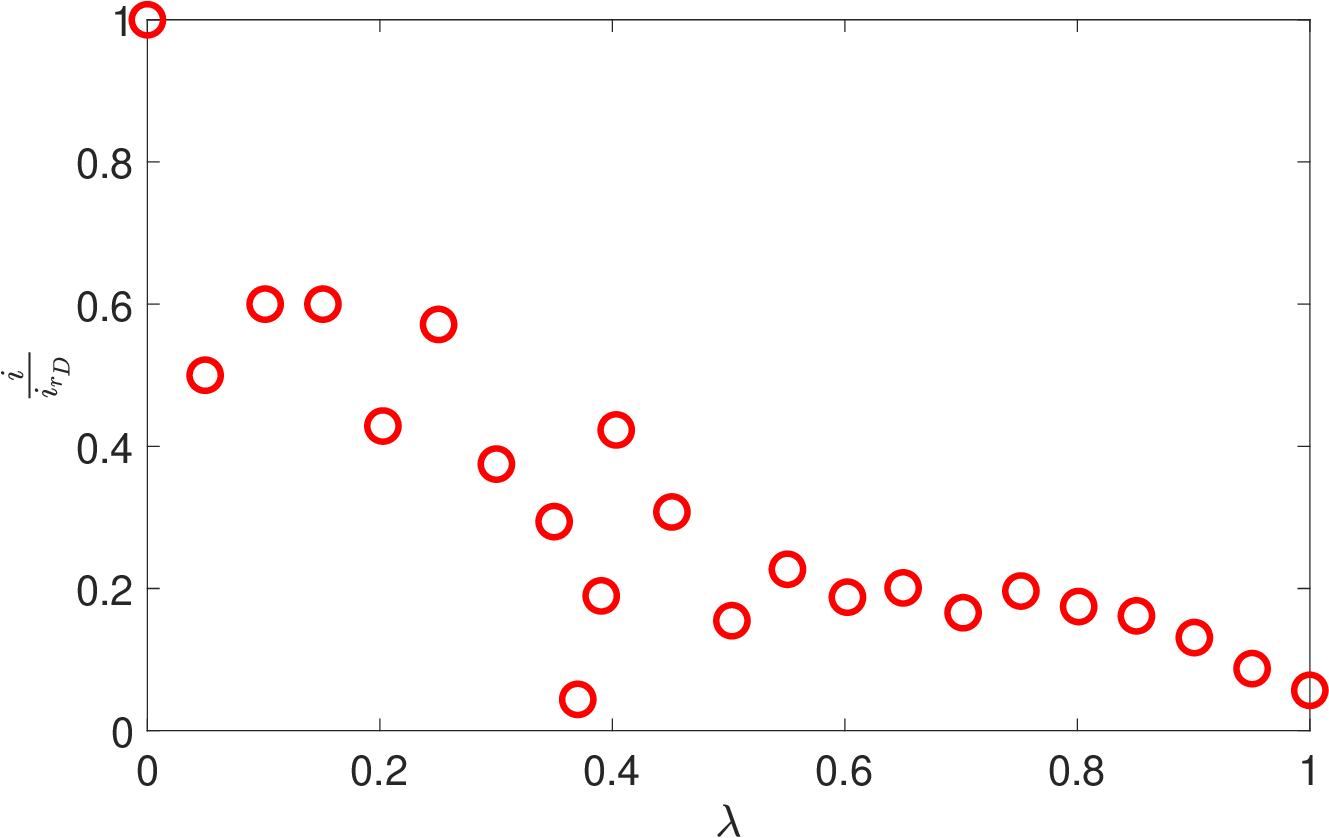}
\caption{Normalized position of knee}\label{f:cellAuto_Norm_pos_knee}\end{subfigure}
\begin{subfigure}[t]{0.3\textwidth}
\includegraphics[width=\textwidth]{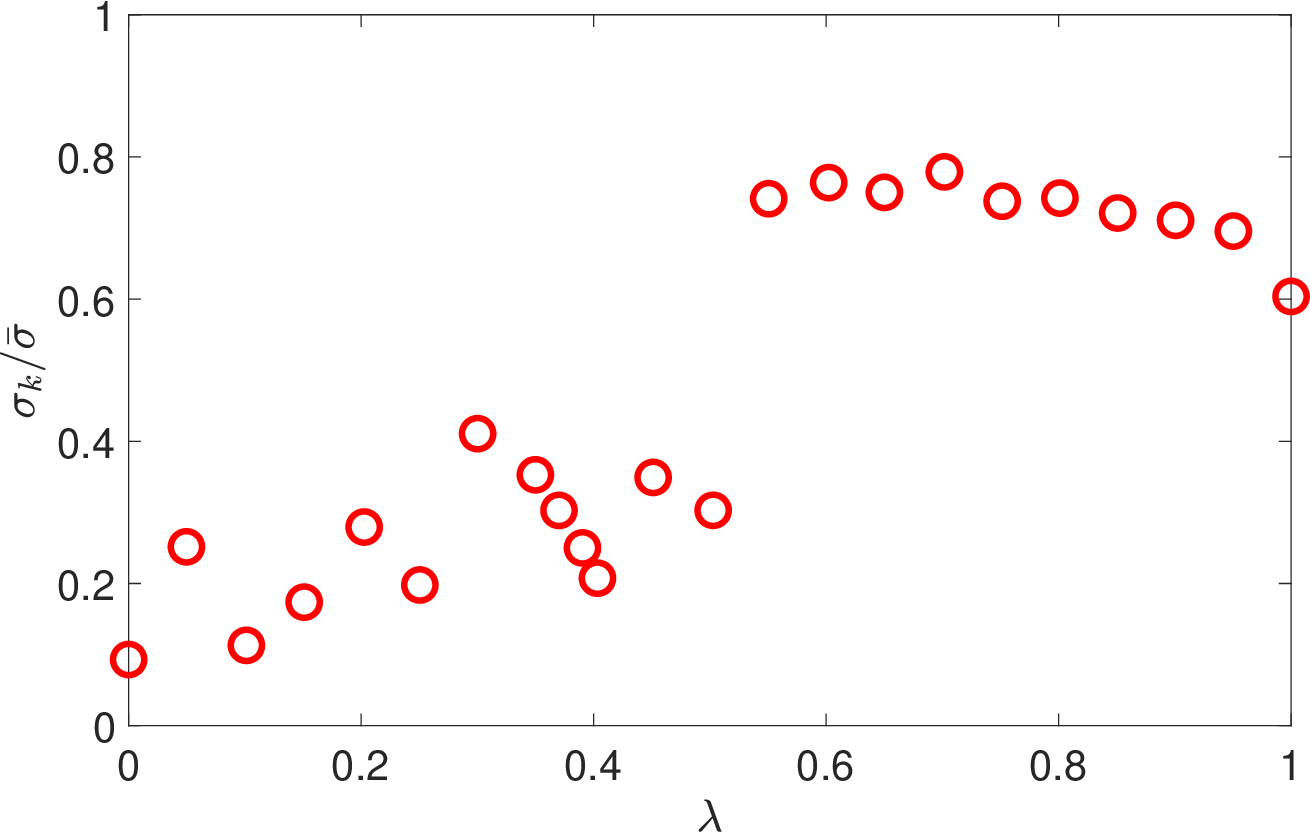}
\caption{Normalized singular value at knee}\label{f:cellAuto_Norm_SV_knee}\end{subfigure}
\begin{subfigure}[t]{0.3\textwidth}
\includegraphics[width=\textwidth]{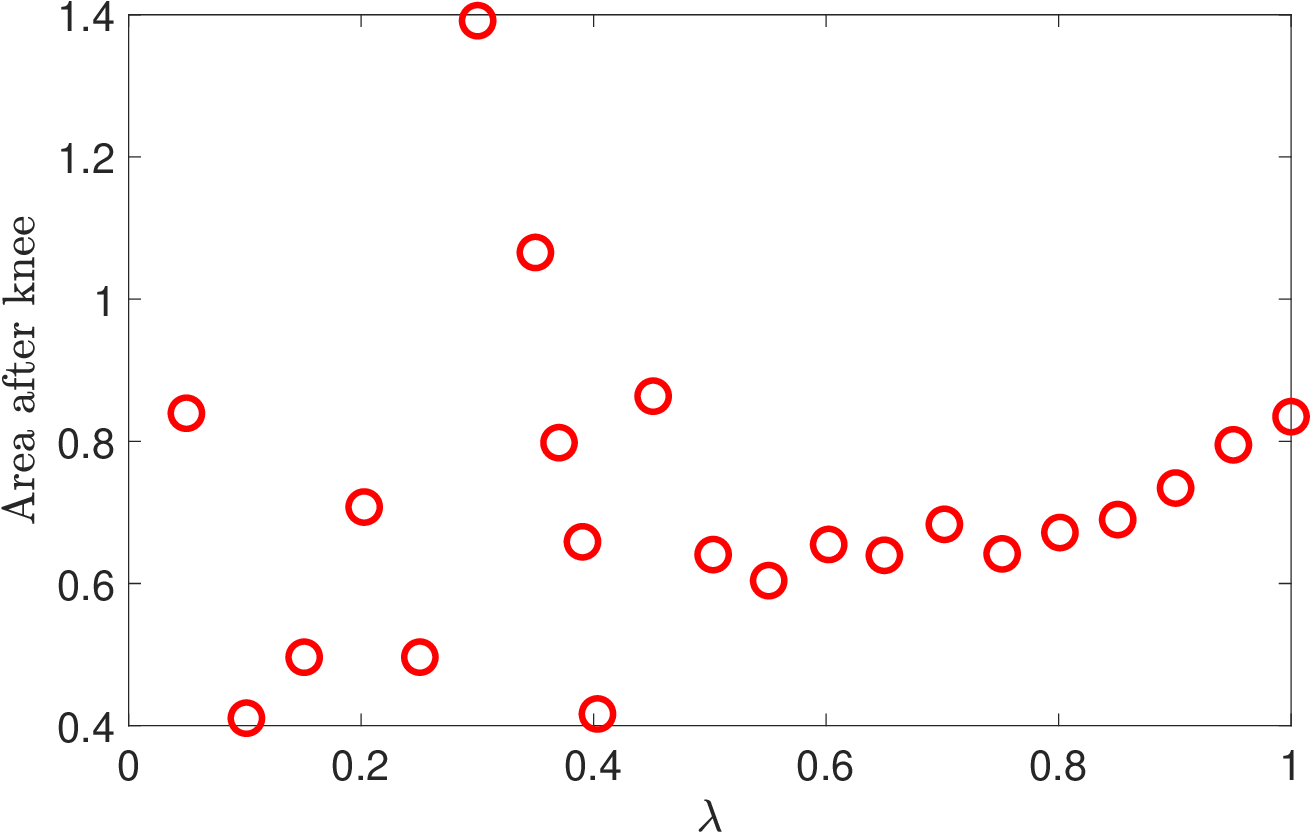}
\caption{Area after knee}\label{f:cellAuto_AreaAfterKnee}\end{subfigure}
\begin{subfigure}[t]{0.3\textwidth}
\includegraphics[width=\textwidth]{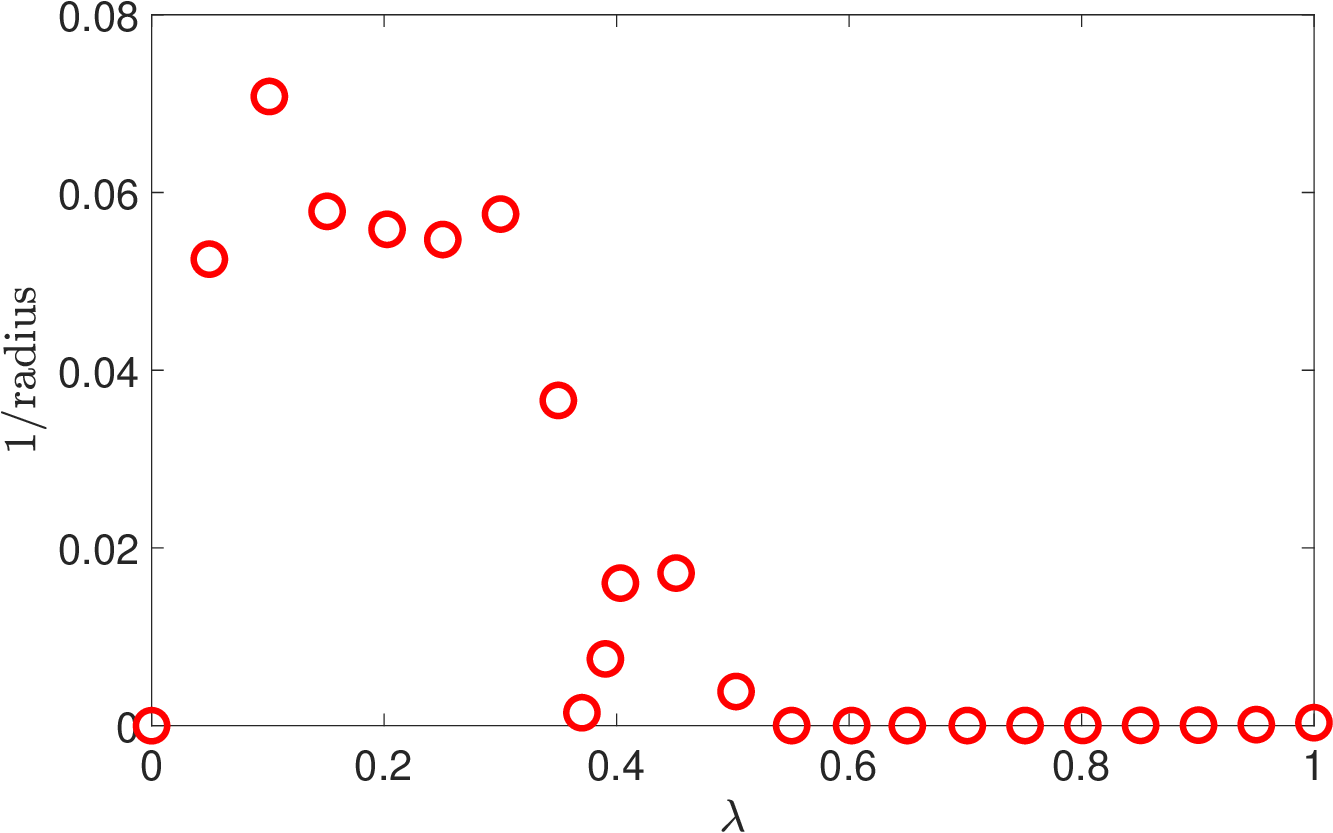}
\caption{Curvature}\label{f:cellAuto_Curvature}\end{subfigure}
\begin{subfigure}[t]{0.3\textwidth}
\includegraphics[width=\textwidth]{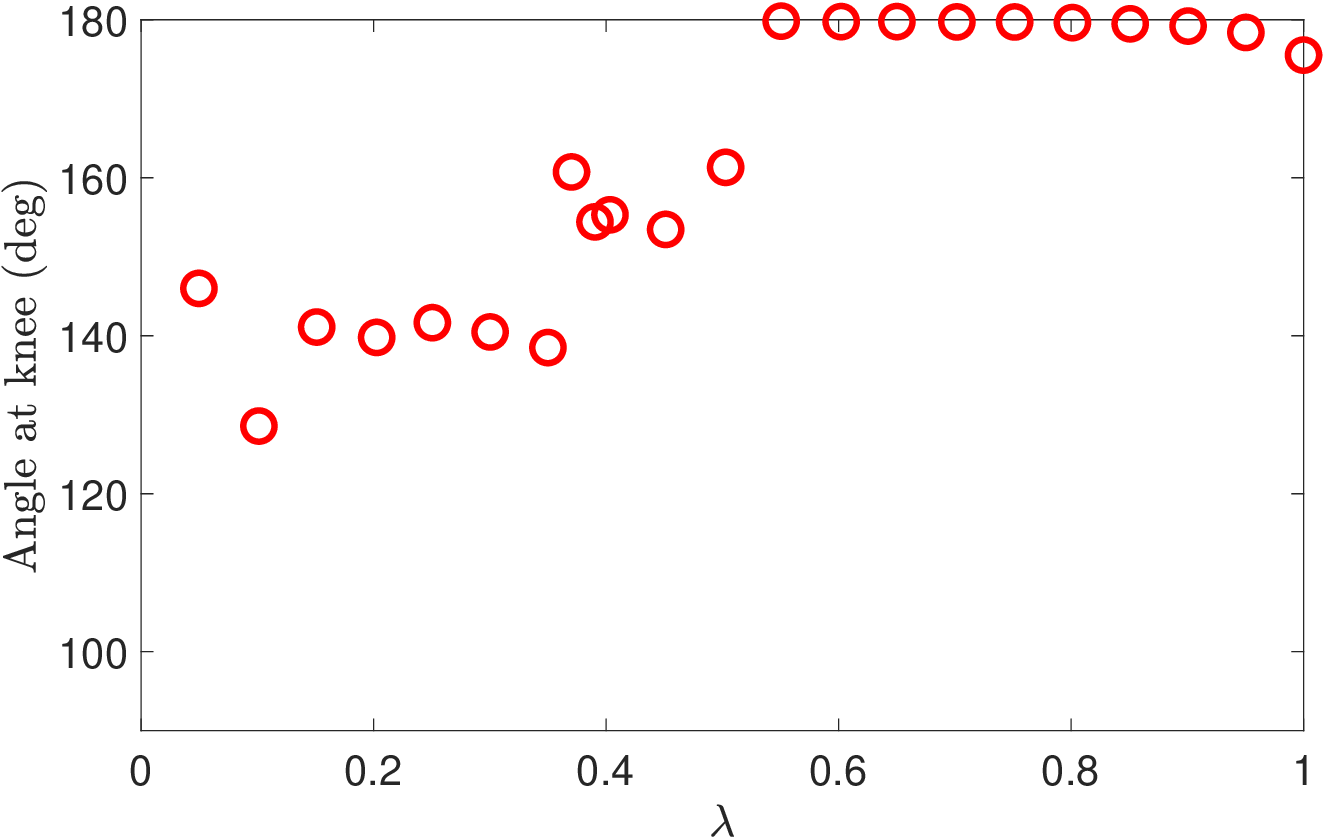}
\caption{Angle at knee}\label{f:cellAuto_AngleAtKnee}\end{subfigure}
\begin{subfigure}[t]{0.3\textwidth}
\includegraphics[width=\textwidth]{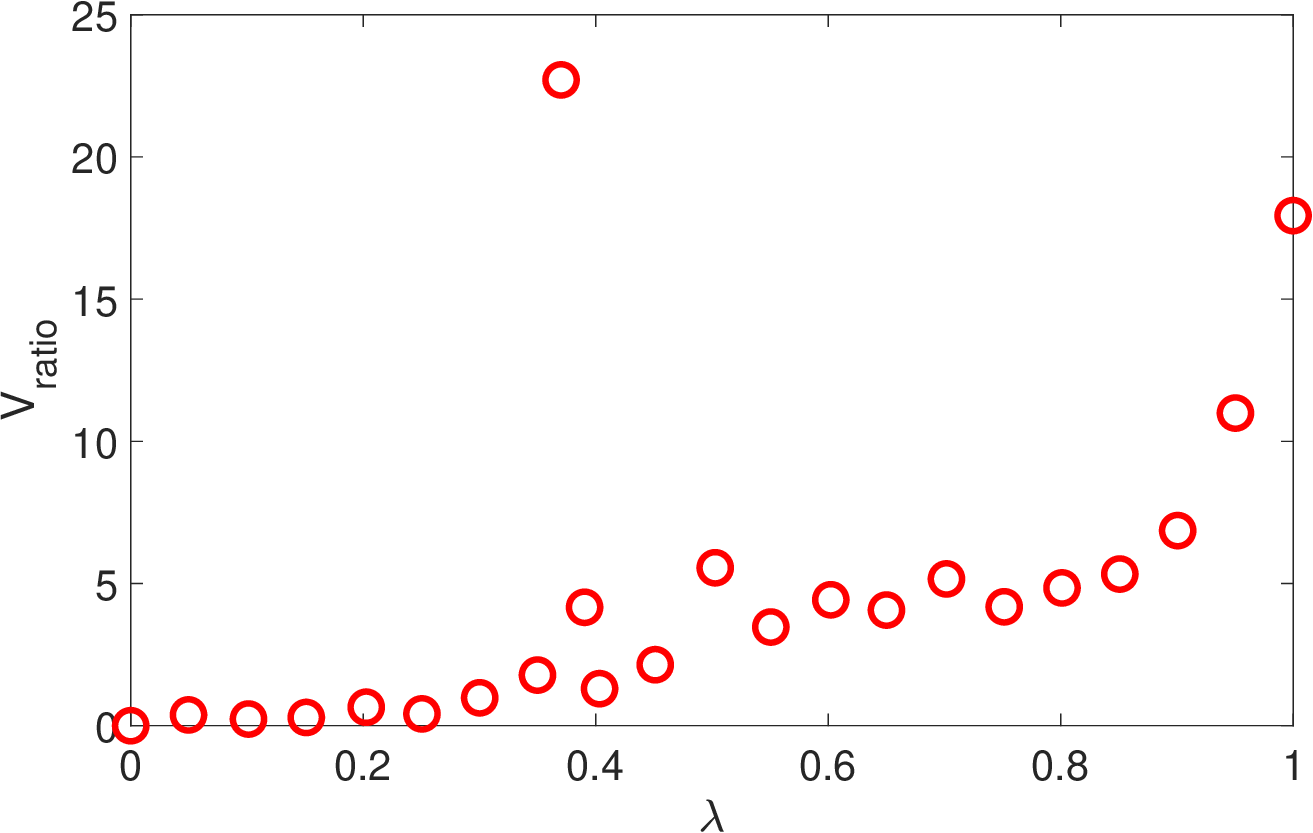}
\caption{Knee vector length ratio}\label{f:cellAuto_Vratio}\end{subfigure}
\caption{Results on Cellular automata tests}
\label{f: cellAutoResults}\end{figure*}

The raw singular value curve in Fig.~\ref{f: cellAutoResults} shows distinct features in the high $\lambda$ curves, including a long sloping region beyond the knee. After entering the ``chaos'' region, the fraction of singular values outside noise bounds drop to zero. The normalized knee position shifts gradually towards higher singular values with increasing $\lambda$. The normalized singular value at the knee increases abruptly while increasing past ``edge of chaos'' region. The knee angle divides the behavior of $\lambda$ in three regions, at low $\lambda$ the angle is close to $140^\circ$, in the edge of chaos region it transitions to $160^\circ$ and in the chaotic region it is nearly $180^\circ$. The knee curvature also gradually drops in the intermittent region, while showing higher variability at low $\lambda$ values that may complicate its use as a detection tool.  Overall, the knee angle's three-phase behavior is most consistent with identifying the underlying $\lambda$ regions used in the simulation

\subsection{SV curve decay discussion} This section discusses the singular value curve behavior across all examples, including the statistical variation in ensemble simulations, theoretical interpretations of some metrics, and comparison to biological data.
\subsubsection{Overall SV decay behavior}
Figure \ref{fig: StatTestAll} illustrates the singular value curves for example motions.   
The singular values curves for pure noise data start marginally higher than $1$ (due to the finite sample size discussed in Section \ref{ss:svDecayNoise}), then decrease almost linearly. This nearly constant gradient decrease is in contrast to more structured motion time histories, where the singular value curve's high and low asymptotic slopes form a knee region.

The constant gradient decline trend is characteristic of pure noise data. The theoretical context is that for infinite sample sizes, noise consists of principal components of equal size, and no sudden drop in singular values is expected. Accordingly, the normalized singular value at knee is higher than all other datasets. 

The combined structure and additive noise datasets show a similar trend on the right side of the knee. After the important information about the dataset (e.g., its motion model) has been captured by the dominant singular values, the remaining singular values carry information about the remaining noise term. The knee is heuristically used to indicate the transition region between dominant and non-dominant singular values. This choice is consistent with multi-objective optimization techniques \citep{Yen2021evoKnee} that select the knee solution among other feasible solutions on the pareto optima. 

The trajectory temporal variance normalization (see Sec.~ \ref{s:skierAlgo})  ensures the results are unaffected by noise of differing variances and provides that singular values of the pure noise trajectories converge to 1 while preserving the observability of inter-signal relationships.  

The singular value curve is insensitive to data matrix row or column ordering and thus indices may be permuted without change to test results, which quantifies a list of visited states rather than an ordered list.  This property does not hold in general for other matrix factorization approaches, e.g. QR decomposition.

\subsubsection{Behavior with respect to noise bounds}
The fraction of singular values outside the noise bound generally behaves as a noise detector based on both swarm motion (Fig.~ \ref{f:225sims_svs_outside}) and cellular automata trajectory (Fig.~ \ref{f:cellAuto_svs_outside}) data. A dominance of singular values remain outside the theoretical noise bound for the more ordered swarms (from acceleration noise through more deterministic motions), a trend which repeats for $\lambda>0.5$ in cellular automata trajectories.

\subsubsection{Knee value, location, and post-knee area}
The normalized singular value at knee in Fig.~\ref{f:225sims_norm_SV_knee} serves as a similar noise detector with a theoretical intuition that will be described in section \ref{sec:RelEckartYoung}. Higher values suggest low compressibility is prominent in noise like swarms and cellular automata trajectories at $\lambda>0.5$. 

The normalized knee position in Fig.~\ref{f:225sims_norm_pos_knee} shows a decreasing trend with decreasing noise levels in swarms that include noise, while noiseless swarm motion simulations show higher values. The greater variability in pure noise motion results may be related to the complexity of defining a knee position for the nearly constant decay seen in noise-related motion curves. For noiseless swarm simulations, the variation may be related to matrix rank variability.  When applied to CA in Fig.~\ref{f:cellAuto_Norm_SV_knee}, normalized knee position shows a consistent decay with greater variability in and below the ``edge of chaos'' region.

Conversely, the area after knee for both swarm simulations in Fig.~\ref{f:225sims_areaAfterKnee} and cellular automata data in Fig.~\ref{f:cellAuto_AreaAfterKnee} shows different behaviors.  In particular, post-knee area reduces from least to most ordered swarm motion model, while this decay is not present in cellular automata, which shows higher variability in and below the edge of chaos region. A metric with embodiment-specific trends has less applicability to the emergence detection focus of this paper and may have more potential for detecting embodiment classes.

\paragraph{Compressibility and relationship to Eckart-Young Theorem}\label{sec:RelEckartYoung}
The classic Eckart-Young theorem \citep{eckartYoung1936svd} provides some context for normalized singular value at knee's behavior differences between more ordered and less ordered motions. This theorem establishes a relation between a matrix norm and its singular values as
\beq\|A-A_i\|_2=\sigma_{i+1},\eeq where $A_i$ is the $i^{th}$ rank approximation of the matrix $A$ and $\sigma_{i+1}$ is its $(i+1)$th largest singular value.
One can then write the normalized knee singular value as
\beq\frac{\sigma_{k}}{\bar\sigma} = \frac{\|A-A_{k - 1}\|_2}{\|A\|_2}\label{eq:normKneeSVmeaning},\eeq
showing that the normalized knee singular value $\frac{\sigma_{k}}{\sigma_1}$ serves as a data compression factor between the truncated and reconstruction  $A_{k -1}$ and the original data contained $A$. This metric indicates how strongly the data set can be compressed without significantly compromising the quality of the dataset. In highly ordered motion models, the interaction structure leads to a more compressible dataset.

\subsubsection{Knee shape and vectors} The knee shape-related metrics considered were curvature, knee angle, and knee vector length ratio. 
In swarm simulations (Fig.~\ref{f:225sims_curvature}), generally, higher curvature represented more ordered motion. In case of cellular automata (Fig.~\ref{f:cellAuto_Curvature}, a gradual decrease in curvature with higher $\lambda$ is seen (with some outliers) and almost zero curvature is seen after $\lambda>0.45$.

In swarm simulations (Fig.~\ref{f:225sims_AngleAtKnee}), pure, kinematic, and acceleration noise had a decaying trend of knee angles corresponding to roughly $180^\circ,150^\circ,120^\circ$. Other more ordered swarms stayed below this level. In case of cellular automata trajectories (Fig.~\ref{f:cellAuto_AngleAtKnee}), roughly three levels at angles $140^\circ,155^\circ,180^\circ$ are seen corresponding to premature, edge-of-chaos, and disordered trajectories.  For swarm trajectories, the singular knee angle shows a gradated decay from pure noise, through the random walks and noise integrations of varying levels, into noisy swarm motion models. The singular knee angle's ability to segment behavioral regions across both swarm motion models and cellular automata trajectories distinguished it as the most applicable metric for trajectory-based order/disorder detection to support emergence recognition across embodiments. Swarm motion models including additive Gaussian noise show a lower knee angle variation, an effect most visible in the Vicsek case where adding noise reduces the knee angle from from 40$^\circ$ variation to 20$^\circ$ variation. Beyond the lower variation seen on noisy examples more consistent with those that may be seen in real-world data. 
This lower variation also provides some background for contemporary swarm simulation construction approaches that add noise to a flock simulation to better match the spectral energy observed in the trajectory probability density functions \citep{wangBSwarmNoise}, a construction that may provide a clearer SKI ORDER signature than a purely deterministic approach.  An exact threshold to determine the edge of chaos behavior across both swarm and cellular automata cases requires more attention due to variation in value and the differing presence of chaos, and observing the trend within similar embodiment classes is useful in discovering ordered/disordered motion.

The knee vector length ratio in Figs.~\ref{f:225sims_Vratio} and  \ref{f:cellAuto_Vratio} shows a generally increasing ratio with $\lambda$ for cellular automata.  In swarm motion models, it generally shows high variability for both noise and deterministic cases, while showing low variability for mixed motion and noise cases, suggesting that the presence of noise may serve to increase the consistency of the knee vector length ratios.

\subsubsection{Comparison to biological example}
\begin{figure*}[ht]    \centering
    \begin{subfigure}[t]{0.3\textwidth} \includegraphics[width=\textwidth]{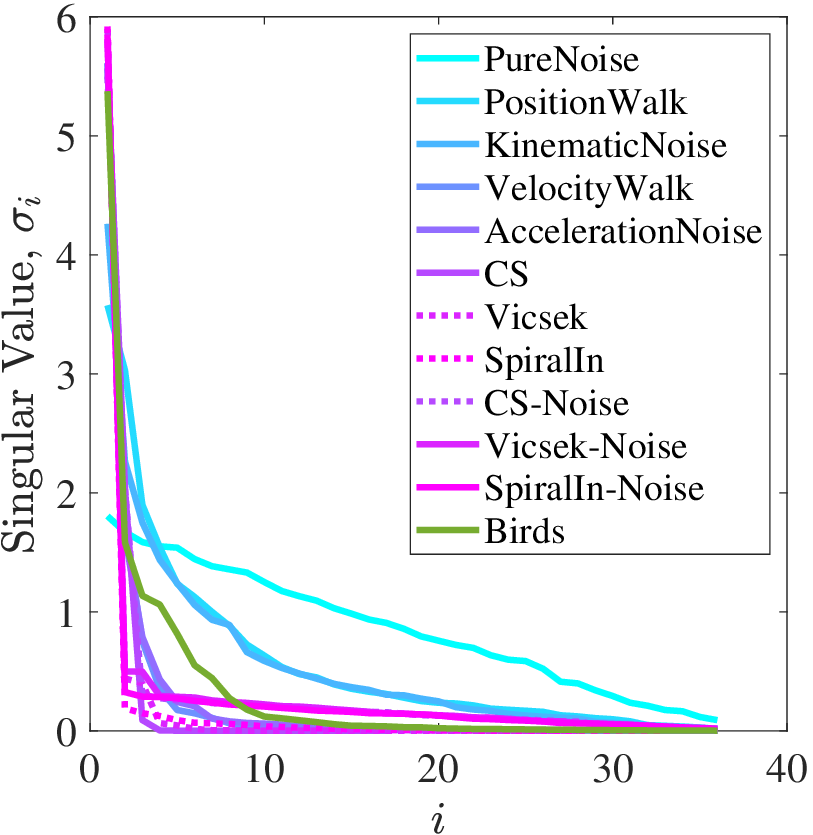}
    \caption{Raw SVD}\label{fig:SV_rawBirdGab}\end{subfigure}    
\begin{subfigure}[t]{0.326\textwidth}
    \includegraphics[width=\linewidth]{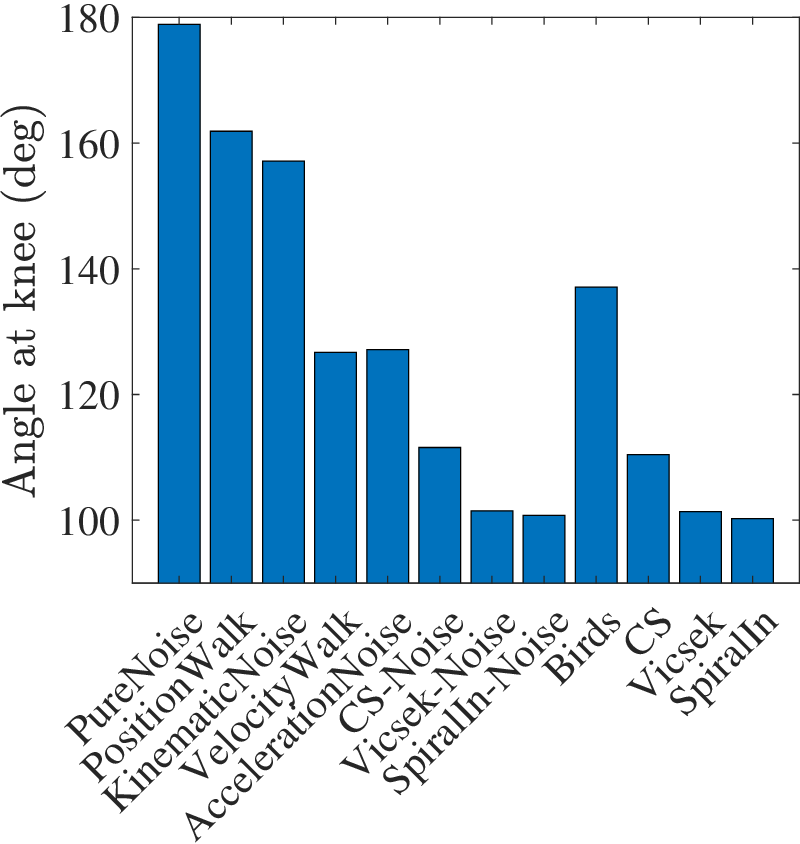}
    \caption{Angle at knee}
    \label{fig:SV_AngleAtKneeBirdgab}\end{subfigure}
\begin{subfigure}[t]{0.317\textwidth} \centering
    \includegraphics[width=\textwidth]{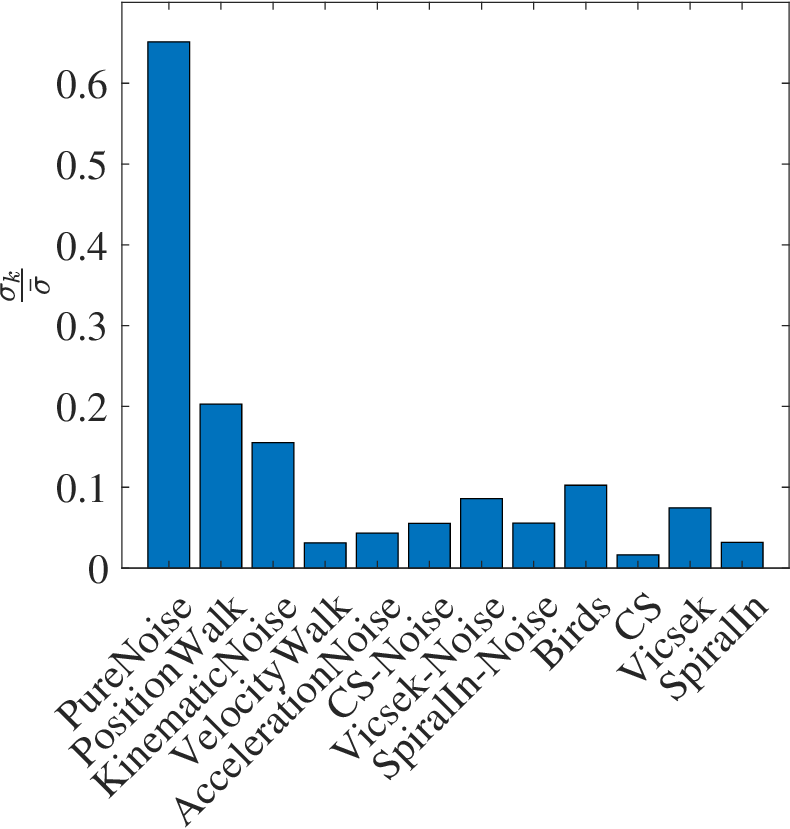}     
   \caption{Value of SV at knee (normalized)}
    \label{fig:SV_valueKneeNormBirdgab} \end{subfigure}    
\begin{subfigure}[t]{0.337\textwidth} 
    \includegraphics[width=\linewidth]{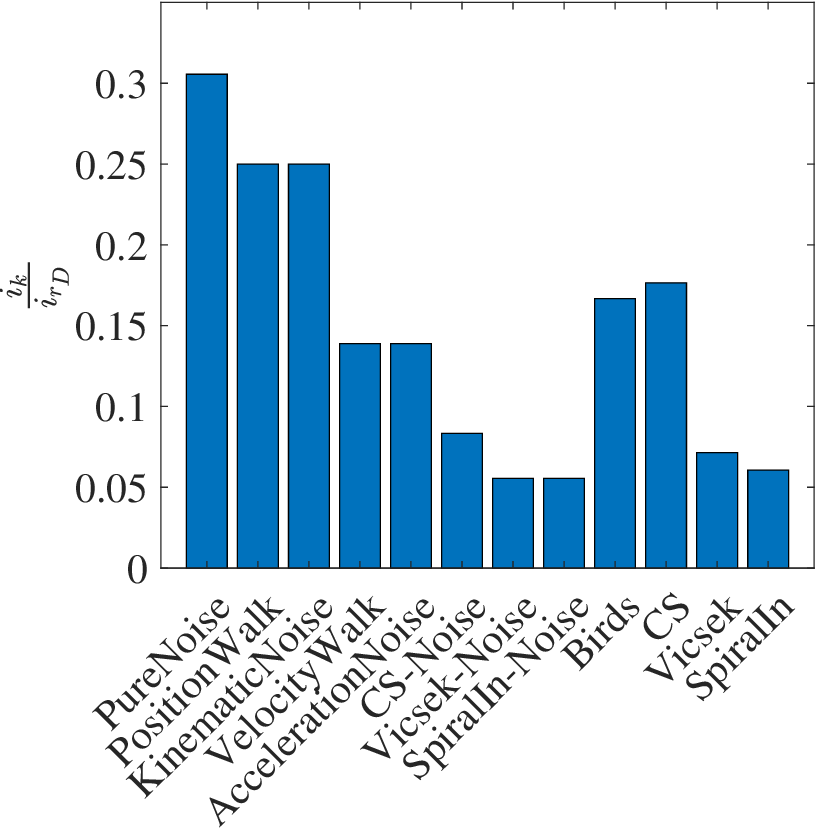}
    \caption{Position of knee (normalized)}
    \label{fig:SV_posKneeNormalBirdgab}\end{subfigure}
\begin{subfigure}[t]{0.326\textwidth}
    \includegraphics[width=\linewidth]{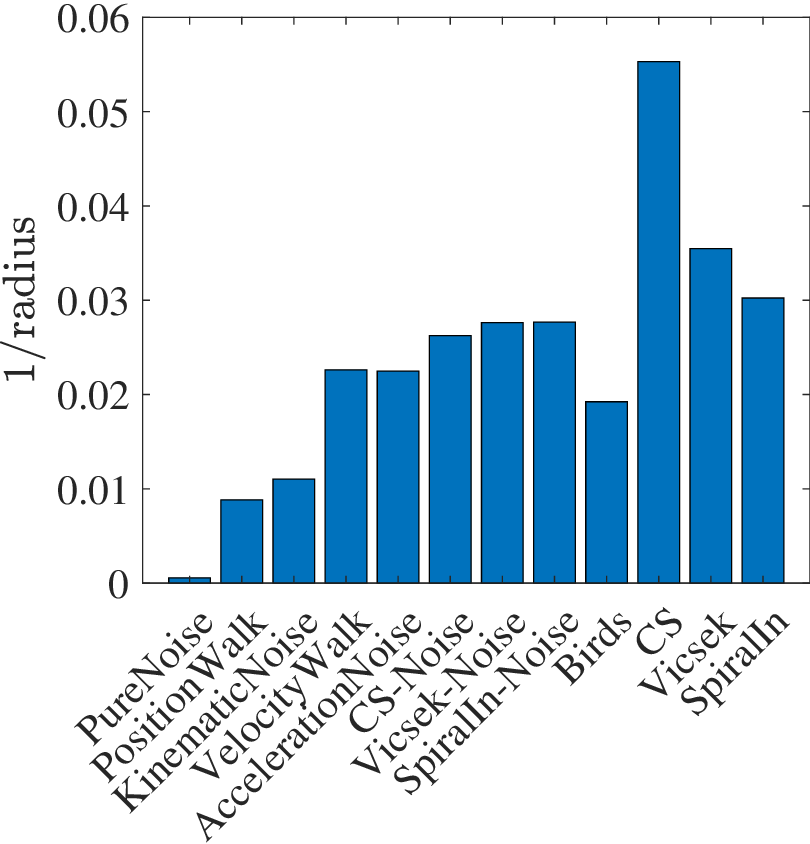}
    \caption{Curvature}
    \label{fig:SV_CurveAtKneeBirdgab}\end{subfigure}
\begin{subfigure}[t]{0.326\textwidth}
    \includegraphics[width=\linewidth]{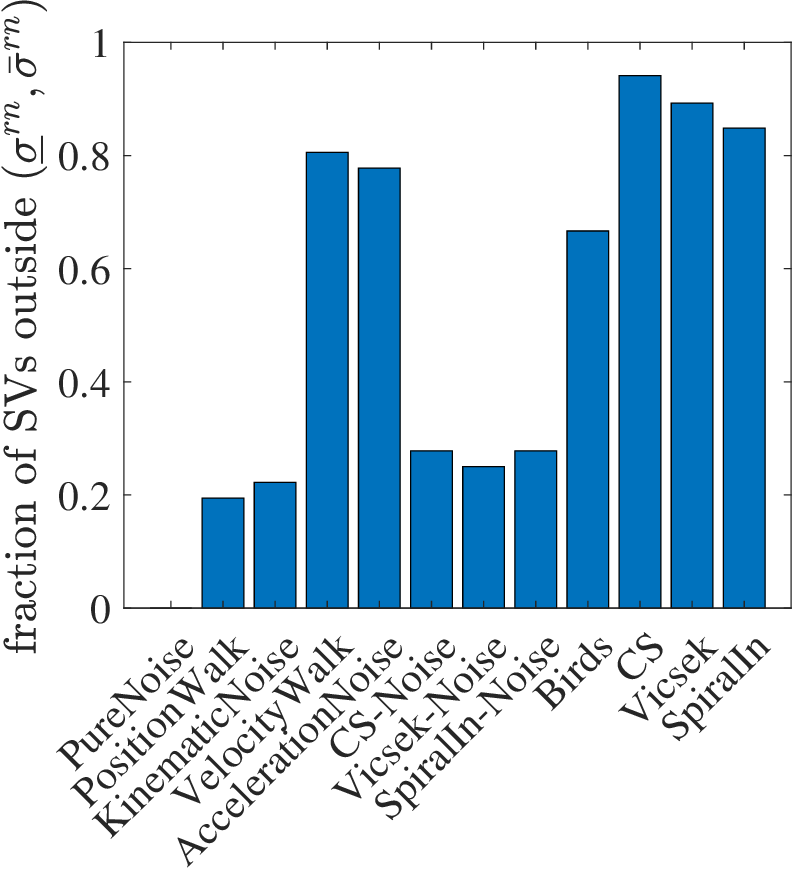}
    \caption{SV's outside random noise bound}
   \label{fig:SV_DominantBirdgab}\end{subfigure}    
    \caption{Singular value curves for example motions, including noise, and a flocking bird video (18 agents 41 frames)}\label{fig:SV_Gab}\end{figure*} 
    
Results of different swarm simulations having the same number of agents and frames with a flocking bird example are presented in Fig.~\ref{fig:SV_Gab}. Only the most significant metrics identified through statistical tests are presented.

The angle at the knee showed a general decreasing trend, indicating increasingly ordered motion. Bird flocks exhibit characteristics that fall between kinematic and acceleration noise, which is further supported by examining the curvature at the knee. Normalized singular values, excluding those for pure noise, position walk, and kinematic noise, are less than $0.1$.

In the cases of acceleration noise, velocity walk, bird flocks, CS, Vicsek, and SpiralIn simulations, over $60\%$ of their singular values lie outside the random noise boundary. Adding a small amount of noise to the simulations makes them resemble pure noise more closely, likely due to the limited number of timesteps.

In this example, the limited number of timesteps results in a relatively high $\kappa = \frac{36}{41} >> 0$. The results would align more closely with the statistical variability results if the simulation datasets had $\kappa = \frac{1}{5}$, which is much closer to zero.

\subsection{Limitations \& areas for improvement}
The singular value decomposition of a data matrix depends on the coordinate system chosen to represent the data. In our swarm examples, time series of positions were used for analyses.  Although any invertible function of position could be used to unambiguously represent the same data, the analysis results could differ. Some robustness to this risk is observed in the flocking bird example where the analyzed video represents a non-invertible projection of the agent positions. The current study's choice of 
coordinates is thus arbitrary. Care must be taken in coordinate choice and data representation as this approach is generalized to other model classes, such as neural network-derived trajectories.  

Singular values are one of the numerous macrostates one could choose for a time-series collection, and this study did not attempt to identify an optimal method of choosing a macrostate, which may require defining a fitness function and searching over possible constructions. 
Rather, SKI ORDER is a constructive approach with theoretic and heuristic support that has the potential to generalize across embodiment classes.

The row-to-column ratio $\kappa$ is used in defining the statistical bounds, thus comparing analysis between two datasets currently requires analyzing the same number of timesteps, and for some metrics, the same number of agents.  Future work is needed to interpret results taken at differing $\kappa$ scales.  Similarly, while the behavior of some metrics may be consistent across examples, the specific values of the quantified metrics may show embodiment class-specificity.  

\section{Summary and conclusions}
Consistent definitions of emergence that provide testable results observable in trajectories are an area with limited results, yet increasing usage of automated intelligence in daily life creates concerns about when a large-scale system may be showing signs of emergence.  Systematic tools for detecting emergence from the observable trajectories across model classes would be a helpful step forward.

This study built on emergence definitions from complexity theory to develop trajectory analysis tools.  The analysis is built on singular value curve analysis of the trajectories and is applied across differing multi-agent motion and cellular automata models.  Eight possible metrics were considered based on statistical noise models and curve behavior.  These metrics and their variability were studied across eleven motion and cellular automata trajectory constructions, as well as compared to biological bird flocking data. Among the metrics, the singular knee angle's ability to segment behavioral regions across both swarm motion models and cellular automata trajectories, and its performance on combined swarm motion and noise models distinguished it as the most promising metric to support emergence detection across embodiments. The fraction of singular values outside noise bounds provides an indication of both the adequate sample size for noise and structure detection and provides noise detection under these criteria. This work represents a foundation to generalize trajectory-based analysis on larger model classes, such as the large-scale partially observable decentralized systems, such as the neural networks applied in artificial intelligence networks, stochastic differential equations, or statistical mechanics.

\FloatBarrier
% \newpage
% \clearpage
\bibliographystyle{abbrvnat}
\bibliography{refs}

\newpage \onecolumn
\section*{Supplementary information: Cellular automata computation}
The \citet{Eck1DCA} cellular automata code used in this study implements the following functions.
\begin{footnotesize}%\twocolumn[]{}
\begin{verbatim}
function newRuleSetData(seed) {
   ruleSeed = seed || Math.floor(Math.pow(2,32)*Math.random());
   random = new Math.seedrandom(ruleSeed);
   var ruleCt = Math.pow(states, neighbors);
   rule = window.Uint8Array? new Uint8Array(ruleCt) : new Array(ruleCt);
   ruleIsUsed = window.Uint8Array? new Uint8Array(ruleCt) : new Array(ruleCt);
   rule[0] = 0;  // always dead
   for (var i = 1; i < ruleCt; i++) {
      ruleIsUsed[i] = 0;
      rule[i] = randInt(1,states-1);
      if (isIsotropic)
         rule[isotropicMate(i)] = rule[i];
   }
   var lambdaCt;
   if (isIsotropic)
      lambdaCt = ((Math.pow(states, neighbors) + Math.pow(states, (neighbors+1)/2)) / 2) - 1;
   else
      lambdaCt = ruleCt - 1;
   lambdaPath = window.Uint32Array? new Uint32Array(lambdaCt) : new Array(lambdaCt);
   var ct = 0;
   for (var i = 1; i < ruleCt; i++) {
      if (!ruleIsUsed[i]) {
         lambdaPath[ct] = i;
         ct++;
         ruleIsUsed[i] = 1;
         if (isIsotropic)
            ruleIsUsed[isotropicMate(i)] = 1;
      }
   }
   for (var i = 0; i < lambdaCt; i++) {
      var r = randInt(0,lambdaCt-1); 
      var temp = lambdaPath[i];
      lambdaPath[i] = lambdaPath[r];
      lambdaPath[r] = temp;
   }
   savedWorlds = null;
   ruleInfo = states + " states, " + neighbors + " neighbors, " +
       (isIsotropic? "isotropic, " : "anisotropic, ") + (lambdaCt+1) + " rules";
   setRulesUsed(0.33*lambdaCt);  // NB: this resets the rulesUsed[] array.
}

function newWorldData(type,seed) {
   worldSeed = seed || Math.floor(Math.pow(2,32)*Math.random());
   worldType = type;
   random = new Math.seedrandom(worldSeed);
   if (savedWorlds == null || (savedWorlds[0] && savedWorlds[0].length != worldSize)) {
       savedWorlds = new Array(worldsInCanvas);
   }
   generationNumber = 0;
   if ( ! savedWorlds[0] ) {
       savedWorlds[0] = window.Uint8Array? new Uint8Array(worldSize) : new Array(worldSize);
   }
   currentWorld = savedWorlds[0];
   var i, start, clumpSize, top;
   if (type > 3) {
       for (i = 0; i < worldSize; i++)
          currentWorld[i] = 0;
   }
   switch (type) {
      case 1:
         for (i = 0; i < worldSize; i++)
            currentWorld[i] = randInt(1,states-1);
         break;
      case 2:
         top = Math.ceil(worldSize/2);
         for (i = 0; i < top; i++)
            currentWorld[i] = currentWorld[worldSize-i-1] = randInt(1,states-1);
         break;
      case 3:
         for (i = 0; i < worldSize; i++) {
            currentWorld[i] = (Math.random() < 0.5)? 0 : randInt(1,states-1);
         }
         break;
      case 4:
         for (i = 0; i < worldSize; i++) {
            currentWorld[i] = (Math.random() < 0.75)? 0 : randInt(1,states-1);
         }
         break;
      case 5:
         clumpSize = Math.min(100,Math.floor(worldSize/3));
         start = Math.floor((worldSize-clumpSize)/2);
         for (i = start; i < start+clumpSize; i++) {
            currentWorld[i] = randInt(1,states-1);
         }
         break;
      case 6:
         clumpSize = Math.min(100,Math.floor(worldSize/3));
         start = Math.floor((worldSize-clumpSize)/2);
         top = Math.ceil(worldSize/2);
         for (i = start; i < top; i++)
            currentWorld[i] = currentWorld[worldSize-i-1] = randInt(1,states-1);
         break;
      case 7:
         clumpSize = Math.min(50, Math.floor(worldSize/5));
         var clumpCt = Math.floor(worldSize/(2*clumpSize));
         start = Math.floor((worldSize - clumpSize*(2*clumpCt-1))/2);
         for (var j = 0; j < clumpCt; j++) {
             for (i = start; i < start+clumpSize; i++) {
                 currentWorld[i] = randInt(1,states-1);
             }
             start += 2*clumpSize;
         }
         break;
      case 8:
         currentWorld[Math.floor(worldSize/2)] = 1;
         break;
      case 9:
         for (i = 4; i < worldSize-2; i+= 10) {
             currentWorld[i] = randInt(1,states-1);
         }
   }
}

function setRulesUsed(used) {
   rulesUsed = Math.round(used);
   if (rulesUsed < 0)
      rulesUsed = 0;
   else if (rulesUsed > lambdaPath.length)
      rulesUsed = lambdaPath.length;
   if (isIsotropic) {
      for (var i = 0; i < rulesUsed; i++) {
         var r = lambdaPath[i];
         ruleIsUsed[r] = 1;
         ruleIsUsed[isotropicMate(r)] = 1;
      }
      for (var i = rulesUsed; i < lambdaPath.length; i++) {
         var r = lambdaPath[i];
         ruleIsUsed[r] = 0;
         ruleIsUsed[isotropicMate(r)] = 0;
      }
   }
   else {
      for (var i = 0; i < rulesUsed; i++)
         ruleIsUsed[lambdaPath[i]] = 1;
      for (var i = rulesUsed; i < lambdaPath.length; i++)
         ruleIsUsed[lambdaPath[i]] = 0;
   }
   var ct = 0;
   for (var i = 0; i < ruleIsUsed.length; i++) {
       ct += ruleIsUsed[i]? 1 : 0;
   }
}
\end{verbatim}\end{footnotesize}

\end{document}